\documentclass[a4paper,fleqn,usenatbib]{mnras}

\usepackage{mathptmx}
\usepackage{gensymb}

\usepackage[T1]{fontenc}
\usepackage{ae,aecompl}

\usepackage{graphicx}	
\usepackage{amsmath}	
\usepackage{amssymb}	
\usepackage{url}
\usepackage{booktabs,caption}
\usepackage[flushleft]{threeparttable}

\newcommand{\fermilat}{\textit{Fermi}-LAT}
\newcommand{\swift}{\textit{Swift}-XRT}
\newcommand{\srcs}{QSO B0218+357}
\newcommand{\srclens}{B0218+357\,G}

\title[MWL study of \srcs ]{Multiwavelength study of the gravitationally lensed blazar \srcs\ between 2016 and 2020.}

\author[V.~A.~Acciari~et.~al.]{\parbox{\textwidth}{\Large{
V.~A.~Acciari$^{1}$,
S.~Ansoldi$^{2}$,
L.~A.~Antonelli$^{3}$,
A.~Arbet Engels$^{4}$,
M.~Artero$^{5}$,
K.~Asano$^{6}$,
D.~Baack$^{7}$,
A.~Babi\'c$^{8}$,
A.~Baquero$^{9}$,
U.~Barres de Almeida$^{10}$
% ...
}
\newline
\emph{\normalsize Affiliations are listed at the end of the paper}
}}

\date{Accepted XXX. Received YYY; in original form ZZZ}

\pubyear{2020}

\begin{document}
\label{firstpage}
\pagerange{\pageref{firstpage}--\pageref{lastpage}}
\maketitle

\clearpage

\begin{abstract}
    We report multiwavelength observations of the gravitationally lensed blazar \srcs{} in 2016-2020.
  Optical, X-ray and GeV flares were detected. 
  The contemporaneous MAGIC observations do not show significant very-high-energy (VHE, $\gtrsim 100$\,GeV) gamma-ray emission.
  The lack of enhancement in radio emission measured by OVRO indicates the multi-zone nature of the emission from this object.
  We constrain the VHE duty cycle of the source to be $<16$ 2014-like flares  per year (95\% confidence). 
  For the first time for this source, a broadband low-state SED is constructed with a deep exposure up to the VHE range. 
  A flux upper limit on the low-state VHE gamma-ray emission of an order of magnitude below that of the 2014 flare is determined. 
  The X-ray data are used to fit the column density of $(8.10\pm 0.93_{\rm stat}) \times 10^{21}\mathrm{cm^{-2}}$ of the dust in the lensing galaxy.  
  VLBI observations show a clear radio core and jet components in both lensed images, yet no significant movement of the components is seen. 
  The radio measurements are used to model the source-lens-observer geometry and determine the magnifications and time delays for both components. 
  The quiescent emission is modeled with the high-energy bump explained as a combination of synchrotron-self-Compton  and external Compton emission from a region located outside of the broad line region.
  The bulk of the low-energy emission is explained as originating from a tens-of-parsecs scale jet.  
\end{abstract}

\begin{keywords}
Gamma rays: galaxies -- Gravitational lensing: strong -- Galaxies: jets -- Radiation mechanisms: non-thermal -- quasars: individual: QSO B0218+357
\end{keywords}



\section{Introduction}

\srcs , also known as S3\,0218+35, is one of only a handful of Flat Spectrum Radio Quasars (FSRQs) detected in very-high-energy (VHE, $\gtrsim 100$\,GeV) gamma-ray emission.
It has a redshift of $z_s=0.944\pm0.002$ \citep{co03, pa17}.
The source showed strong variability in the GeV range in 2012  \citep{ch14} when a series of flares was observed by \textit{Fermi} Large Area Telescope (LAT).
Another flare was observed by \fermilat\ in 2014, and during the follow-up the source was discovered in VHE gamma rays by MAGIC telescopes \citep{2015arXiv150203134B, 2015ICRC...34..877B, ah16}. 
Similarly to \srcs , GeV emission from the second gravitationally-lensed source  detected by \fermilat{}, PKS\,1830-211, also shows evidence of a measured delay between different lens images \citep{ba11}.  
Observations of PKS\,1830-211 by the H.E.S.S. telescopes following a GeV flare did not show any significant gamma-ray emission \citep{ab19}.

\srcs\ is gravitationally lensed by \srclens , a spiral galaxy seen face-on at a redshift of $z_{l}=0.68466\pm0.00004$ \citep{cry93}.
Strong gravitational lensing is observed when the lens is a galaxy or a cluster of galaxies. Such a massive lens can produce multiple images of the source separated by  $\sim$ arcseconds. Thus, the images of strongly lensed sources can be well resolved at wavelengths from radio to X-rays with existing instruments. 

Stars can cause further lensing effects within a lensing galaxy. In such cases, the deflection angle of lensed images is of the order of microarcseconds. Thus, the effect is called microlensing. The change in position of microlensed images cannot be observed with existing instruments. The microlensing effect is observed as changes in the flux of the strongly lensed image. 

The relative flux densities observed for lensed images depend on the geometry of the source-lens-observer system, and can be affected by microlensing.
Further, different geometrical paths and gravitational delays cause the emission to arrive at different times in various images.
In the case of \srcs\ the image is composed of two images A and B separated by only 335\,mas  and an Einstein ring of a similar size \citep{od92}. 
The A component (located westwards) is brighter and this signal precedes that from component B.

Variable radio emission observed in 1992/1993 and 1996/1997 with the Very Large Array at 5, 8.4 and 15 GHz yields time delays between these two components of
$12\pm3$\,d \citep{co96}, 
$10.5\pm0.4$\,d \citep{bi99}, 
$10.1^{+1.5}_{-1.6}$\,d, \citep{co00}, 
$11.8\pm2.3$\,d \citep{em11}, 
$11.3\pm0.2$\,d \citep{bb18}. 
The statistical analysis of the 2012 high state \fermilat\ $>0.1$\,GeV light curve auto-correlation function led to a similar value of the time delay ($11.46\pm0.16$\,d, \citealp{ch14}).
These values are consistent with the delay between the two components of the 2014 \fermilat\ flare \citep{ah16}.

The gamma-ray emission of \srcs\ comprises many flares with timescales as short as a few hours. The short timescales of gamma-ray flares combined with the ability of the \fermilat\ observatory to monitor the sky continuously allow one to search for delayed counterparts of flares and put constraints on the magnification ratio. For example, the two-night-long sub-TeV flare was observed contemporaneously with the detection of the image B flare in \fermilat\ \citep{ah16}

Unfortunately, since the MAGIC observations in 2014 only covered the time of the B image of the flare, no measurement of the magnification ratio or delay could be obtained. 
Monitoring of \srcs\ with Cherenkov telescopes during flaring events could enable the capture of multiple flares and constrain models of the magnification ratio and time delays.

At radio frequencies the B component is 3.57-3.73 times fainter than the A component \citep{bi99}. 
However, the observed ratio of magnification varies with the radio frequency \citep{mi06}, possibly due to free-free absorption in the lens \citep{mi07}. 
In the optical range the leading image is strongly absorbed, inverting the brightness ratio of the two images \citep{fa99}.
It has been suggested that the optical absorption occurs in the host galaxy rather than the lens \citep{fa17}. 
Interestingly, the magnification ratio observed at GeV energies shows variability.
The average GeV magnification factor during 2012 high state was estimated to be $\sim1$ \citep{ch14}, while during the 2014 flare it was comparable to or even larger than that  measured at radio frequencies \citep{ah16}.
Changes in the observed GeV magnification ratio can be interpreted as microlensing effects either due to individual stars  \citep{vn15} or due to larger scale structures in the lens \citep{sb16}. 

Because it takes about 1-2 days for \fermilat{} data to be collected, downlinked and processed, it is difficult to trigger observations for phenomena with similar durations, like the two-night 2014 flare.
Therefore, the shortness of the VHE gamma-ray emission significantly hinders the possibility of Target of Opportunity observations of a flare in both images.
In addition observational visibility constraints further limit the possibility of a follow up of the delayed emission with ground based instruments. 
Thus, since 2016, we have taken advantage of the 11 days delay to trigger MAGIC.
Observational windows which allow visibility under favourable zenith angle conditions in moon-less nights 11 days after each slot have been identified.
During these time slots MAGIC observations were performed, and contemporaneous multiwavelength (MWL) coverage from radio to GeV gamma-rays was obtained.
In this paper the results of these observations are reported.
Additionally, a multiwavelength campaign on \srcs{}, organized in August 2020 in response to a hint of enhanced activity in the source, is also reported. 

In Section~\ref{sec:obs} the instruments that took part in the MWL campaign, the data taken and the analysis methods are described.
The results are reported in Section~\ref{sec:res}.
In Section~\ref{sec:model}, the broadband emission of the low state of the source is modeled.
The paper concludes with a summary of the results in Section~\ref{sec:conc}. 

\section{Observations and data analysis} \label{sec:obs}

\srcs\ was observed over a broad energy range:
radio (The Owens Valley Radio Observatory, OVRO), radio interferometry (a joint VLBI array of KVN (Korean VLBI Network) and VERA (VLBI Exploration of Radio Astrometry), KaVA),
optical (Kunliga Vetenskapsakademi, KVA and Nordic Optical telescope, NOT;
\textit{Neil Gehrels Swift observatory (Swift)} Ultraviolet/Optical Telescope
(\textit{Swift}-UVOT) and \textit{X-ray Multi-Mirror} Optical Monitor (\textit{XMM}-OM)), X-ray (X-ray Telescope (\swift) and \textit{XMM}-Newton),
GeV gamma rays (\fermilat )
and VHE gamma rays (MAGIC). 
During the August 2020 MWL campaign dedicated observations by Himalayan Chandra Telescope (HCT), Joan Oró Telescope (TJO) and Mets\"ahovi were taken. 

The historical data obtained via the Space Science Data Center\footnote{SSDC, \url{http://www.ssdc.asi.it/}} from the following catalogs are also used:
CLASS \citep{ssdc_class},
JVASPOL \citep{ssdc_jvaspol},
KUEHR \citep{ssdc_kuehr},
NIEPPOCAT \citep{ssdc_nieppo},
NVSS \citep{ssdc_nvss},
Planck \citep{ssdc_ercsc, ssdc_pccs1, ssdc_pccs2}, 
GB6 \citep{ssdc_gb6},
GB87CAT \citep{ssdc_gb87},
WMAP5 \citep{ssdc_wmap5},
WISE \citep{ssdc_allwise},
1SWXRT \citep{ssdc_1swxrt},
1SXPS \citep{ssdc_1sxps},
FGL \citep{ssdc_1fgl, ssdc_2fgl, ssdc_3fgl}.

\subsection{MAGIC}

MAGIC is a system of two imaging atmospheric Cherenkov telescopes with a mirror dish diameter of 17\,m each. 
The telescopes are located in the Canary Islands, on  La Palma ($28.7^\circ$\,N, $17.9^\circ$\,W), at a height of 2200 m above sea level \citep{al16a}. 
The data were analyzed using MARS, the standard analysis package of MAGIC \citep{za13, al16b}.
Wherever applicable, upper limits on the flux were computed following the approach of \cite{ro05} using a 95\% confidence level and assuming a 30\% total systematic uncertainty on the collection area. 

The regular monitoring observations were performed between MJD 57397 and 58875 in dark night conditions. 
The monitoring time slots were scheduled so as to allow for possible observations in $\sim 11$\, days if enhanced emission was seen. 
This results in possible observation slots (up to two per moon period) lasting between 1 and 6 days. 
Motivated by the 2-day duration of the 2014 VHE flare, in such slots observations every second night were scheduled (on some occasions this scheme was modified due to weather conditions or competing sources). 
After the data selection, based mainly on the atmospheric transmission measured with LIDAR \citep{fg15} and on hadronic background rates,
the data set consists of $72.7$\,hr, spread over 73 nights.

Since MJD 58122 the data have been taken with the novel Sum-Trigger-II \citep{da21}. 
The Sum-Trigger-II part of the dataset consists of 38.4\,hr and was analyzed with dedicated low-energy analysis procedures including a special image cleaning procedure \citep{sh13,ce19}.

Additionally, during the August 2020 campaign, 2\,hr of good quality data were taken on MJD 59081 and 59082. 
Due to a forest fire observations on MJD 59083 could not be used. 
\subsection{\fermilat}

The LAT is a pair conversion detector on the \textit{Fermi} Gamma-ray Space Telescope, which was launched on June 11, 2008. It observes the whole sky every three hours in the energy range between a few tens of MeV and few TeV \citep{fermipaper}.
The \fermilat\ data taken between MJD 56929  
and 58876 
in the energy range 100 MeV -- 2 TeV in a region of interest of 15\degree\ were selected.
The data were processed using the Fermitools version 1.2.23 and Fermipy \citep{fermipy} version 0.19.0, with instrument response function P8R3\_SOURCE\_V2. The data were binned in 8 energy bins per decade and in spatial bins of 0.1\degree. To reduce the contamination from the Earth limb, a zenith angle cut of 90\degree\ was applied to the data.
The model used in the likelihood analysis is composed of the sources listed in the LAT 8-year Source Catalog (4FGL, \citealp{4fgl}) that are within 20\degree\ of the \srcs\ location, the latest interstellar emission model (gll\_iem\_v07), and an isotropic background model (iso\_P8R3\_SOURCE\_V2\_v1). 
At the beginning of the analysis, we iteratively optimized our spectral source models using fermipy's optimization method.
Sources with a Test Statistic\footnote{The Test Statistic is defined as $TS=-2ln(L_{max,0}/L_{max,1})$, where $L_{max,0}$ is the maximum likelihood value for a model without an additional source and $L_{max,1}$ is the maximum likelihood value for a model with the additional source. It is a measure of the detection significance of a source.} (TS) lower than 1 were removed from the fit. Four new point sources with a TS higher than 16 ($\sim 4 \sigma$ significance) within 10$^\circ$ of \srcs\ were added iteratively, in order to account for emission not modeled by known background sources ($\mathrm{RA_{J2000}}$, $\mathrm{Dec_{J2000}}$=30.54\degree, 39.67\degree; 35.55\degree, 37.53\degree \footnote{4FGL J0221.8+3730 is a new source in the  LAT 10-year Source Catalog (4FGL-DR2 \url{https://fermi.gsfc.nasa.gov/ssc/data/access/lat/10yr_catalog/}) compatible with this location.}; 31.72\degree, 38.56\degree; 43.58\degree, 33.63\degree). For each of these sources a power-law spectral model was used and iteratively optimized. The closest new source is 1.6$^\circ$ away from \srcs{}, and has a TS slightly above 40.
The effect of energy dispersion\footnote{\url{https://fermi.gsfc.nasa.gov/ssc/data/analysis/documentation/Pass8_edisp_usage.html}} (reconstructed event energy differing from the true energy of the incoming photon) is accounted for by
 generating a detector response matrix with two additional energy bins in log$E_{true}$ above and below the considered energy range (edisp\_bins = -2). This method is applied to all the sources in the model except for the isotropic background which was derived from dispersed data. The normalisation of both diffuse components in the source model were allowed to vary during the spectral fitting procedure. In the whole interval analysis, sources within 7\degree\ from \srcs\  had their normalisation free to vary, sources within 5\degree\ from \srcs\ had also their spectral index free to vary. In both cases this selection was applied only to sources with a TS in the full time interval MJD 56929-58876 higher than 10. 
The \srcs\ was modeled with a LogParabolic spectrum:
\begin{equation}
 \frac{dN}{dE} = N_0\left(\frac{E}{E_b}\right)^{-(\alpha + \beta\log(E/E_b))} 
\end{equation}
as in the 4FGL.
In the overall analysis, the source was observed with a TS of 7678 ( $\sim 87 \sigma$) with a flux of $(9.70\pm0.31)\times 10^{-8} \mathrm{cm^{-2}\,s^{-1}}$ above 0.1\,GeV. 
The spectral energy distribution (SED) was also evaluated over the whole time range, and over only the days in which \srcs\ was observed by MAGIC. In the second case the summed likelihood technique was used for combining the analysis in the different time bins.

In order to estimate weekly and daily fluxes of \srcs\ the number of free spectral parameters is limited. The sources located within 7\degree\ from \srcs\  had their normalisation set as a free parameter if their variability index was higher than 18.483\footnote{The level was chosen according to the 4FGL catalogue.}, while all sources within  5\degree\ from \srcs\ had their normalization free to vary if their TS was higher than 40 integrated over the full time period. The spectral indices of all the sources with free normalisation were left as a free parameter if the source showed a TS value higher than 25 over the integration time (weekly or daily), in all the other cases the indexes where frozen to the value obtained in the overall fit.

The spectral analysis was also performed in two different smaller intervals corresponding to the optical/GeV flare (MJD 57600-57700) and to the X-ray flare  (MJD 58860.7 -- 58866.7). 
The source was significantly detected in both time intervals, with a significance of $45.9 \sigma$ and $6.6 \sigma$, respectively. 

\subsection{XMM-Newton}

XMM-Newton \citep{jansen01} observed the source four times between August 2019 and January 2020.  The integration times of the observations were in the range of 11.3 ks and 19.8 ks.
The EPIC pn CCD camera \citep{stru01} operated in full-frame mode with the  medium filter applied during all the observations.
The data were processed using the XMM-Newton Science
Analysis System \citep[{\tt SAS} v.18.0.0,][]{gabriel04}
 following standard settings and using the calibration files available at the time of the data reduction. The EPIC pn Observation Data Files
(ODFs) were processed with the  {\tt SAS}-task {\tt epproc}  in order to generate the event files.
Event files were cleaned of bad pixels, and events spread at most in two contiguous pixels
(PATTERN$\leq$4) were selected.
Periods of high background levels were removed by analysing the light curves of the count rate at energies higher than 10 keV. The resulting net-exposure times are reported in Table \ref{XMM_fit}.
In order to include all of the source counts and simultaneously minimise the background contribution, source counts were extracted from a circular region of radius between 30 and 35 arcsec. 
The background counts were extracted from a circular region of radius 50-65 arcsec located on a blank area of the detector close to the source.
Response matrices for spectral fitting were obtained using the  {\tt SAS}-task {\tt rmfgen} and  {\tt arfgen}.
All the spectra were binned in order to have no less than 20 counts in each background-subtracted spectral channel, and the instrumental energy resolution was not oversampled by a factor larger than 3.

X-ray spectral analyses were carried out with XSPEC v.12.9.1 \citep{arnaud96}. No variability in the spectra of the XMM-Newton observations performed  at MJD 58697, 58721, and 58724 (obs. ID 0850400301, 0850400401, 0850400501) was observed, thus all the observations were  combined with the  {\tt SAS}-task {\tt epicspeccombine} for the spectral modelling of the source. In contrast, the observation performed at MJD  58863.7 (obs. ID 0850400601) indicated an increase of the X-ray flux by a factor of $\sim$1.4 with respect to previous observations, thus this spectrum was fitted separately.

\subsection{\swift}
\label{sec:swift-data}

The X-ray Telescope \citep[XRT,][]{2004SPIE.5165..201B} on-board the \textit{Neil Gehrels Swift observatory (Swift)} observed the source four times between January 2016 and January 2020. 
Additionally seven pointings were taken around the time of the August 2020 campaign. 
Due to the source faintness, all of these observations were performed in photon counting mode. The event lists for the period of interest were downloaded from the publicly available SWIFTXRLOG (\textit{Swift}-XRT Instrument Log)\footnote{\url{https://heasarc.gsfc.nasa.gov/W3Browse/swift/swiftxrlog.html}}. The data were processed using the standard data analysis procedure \citep{2009MNRAS.397.1177E} and the configuration described by \citet{2017A&A...608A..68F} for blazars. The spectra of each observation were binned in a way that each bin contains one count.  
Therefore, the maximum likelihood-based statistic for Poisson data \citep[Cash statistics,][]{1979ApJ...228..939C} method was used in the spectral fitting procedure and flux measurements of individual observations.

No spectral variability was observed within the \swift\ data. In order to evaluate the average state of the source during the monitoring, two combined spectra were produced using the observational data taken during 2016-2017 (OBSIDs 00032533003, 00032533005, 00032533006, and 00032533007) and August 2020 (OBSIDs 00032533008, 00032533009, 00032533010, 00032533011, 00032533012, 00032533013, 00032533014, 00032533015). 
These spectra are binned in a way that each bin contains 20 counts. Therefore,  the maximum likelihood-based statistic for Gaussian data method is the spectral fitting procedure of these two spectra. 

\subsection{UV}
During the four monitoring {\em Swift} pointings, the UVOT instrument observed the source in the $u$ optical photometric band \citep{poole08,breeveld10}. The data were analysed using the \texttt{uvotimsum} and \texttt{uvotsource} tasks included in the \texttt{HEAsoft} package (v6.28) with  the  20201026  release  of  the  Swift/UVOTA  CALDB. Source counts were extracted from a circular region of 5 arcsec radius centered on the source, while background counts were derived from a circular region of 20 arcsec radius in a nearby source-free region. 
The source was not detected with a significance higher than 3 $\sigma$ in the single observations, therefore the four UVOT images were summed using the \texttt{uvotimsum} task and analyzed the summed image with the \texttt{uvotsource} task. 

The Optical Monitor (OM) observed the source four times in $u$ filters in imaging mode. The total exposure times of the imaging observations were: 16400 s, 17700 s, 11300 s, and 19800 s. The data were processed using the SAS task \texttt{omichain}. 
For the count rate to flux conversion  the conversion factors given in the SAS watchout dedicated page\footnote{https://www.cosmos.esa.int/web/xmm-newton/sas-watchout-uvflux.} were used .

During the August 2020 campaign five observations with \textit{Swift}-UVOT were taken in $u$ band. 
None of these pointings resulted in the detection of a signal above $2\sigma$ significance. 
Similarly to 2016--2020 monitoring data, a stacked analysis was performed  to evaluate the average emission in this period. 

The UVOT and OM flux densities were corrected for Galactic extinction using a value $\mathrm{A_U}$ = 0.299 mag \citep{sch11}.

\subsection{Optical}

The source was monitored with the 
NOT and 35cm Celestron telescope attached to the 
KVA telescope. Both telescopes are located at the same site as the MAGIC telescopes and the NOT observations were carried out quasi-simultaneously with MAGIC, while KVA performed additional monitoring more frequently. Starting in July 2020 the source was also monitored with the 
TJO at the Montsec Astronomical Observatory\footnote{\url{http://www.ieec.cat/en/content/210/telescope-and-dome}}. 
In addition during the August MWL campaign the source was observed with the 
HCT {\footnote{\url{https://www.iiap.res.in/iao/2mtel.html}}}.  

NOT observations were carried out using ALFOSC in B, V, R and I-bands, while the KVA and TJO observations operated in R-band only. The data were analyzed using the semi-automatic pipeline and standard procedures of differential photometry \citep{nilsson18}. The same comparison and control stars were used as in \citet{ah16}{\footnote{The stars are marked in finding chart available at: \url{http://users.utu.fi/kani/1m/finding_charts/B2_0218+35_map.html}}}.
The r-,g- and i-band magnitudes of the stars were available in the PANSTARRS database\footnote{\url{https://catalogs.mast.stsci.edu/panstarrs/}}. The magnitudes in B, V and I-band were calculated from i-, r- and g- band magnitudes using the formulae of Lupton (2005) \footnote{\url{http://classic.sdss.org/dr4/algorithms/sdssUBVRITransform.html\#Lupton2005}}.
Using those formula, consistency of the R-band values derived in \citet{ah16} was checked. The I-band filter used at the NOT differs from the standard I-band filter enough for the color correction to become significant for a very red input spectrum as in this case \citep{fa17}. The spectrum obtained by \cite{fa17} was downloaded from the ZBLLLAC repository\footnote{\url{https://web.oapd.inaf.it/zbllac/}} and synthetic photometry was performed through the standard I-band and the NOT I-band. This showed that the NOT I-band magnitudes needed to be corrected by +0.08 mag to transform to the standard system.  
The aperture used was 3 arcsec, slightly smaller than the 4 arcsec used for KVA and TJO data. 

The source was observed with HCT in four epochs (MJD 59081--59085). 
The observations were carried out in the Bessell U, B, V, R and I bands available with HFOSC.
The data were reduced in a standard manner using various tasks available in IRAF.
Aperture photometry was performed on the source and  nearby stars.  
The standard magnitude of the source was obtained using differential photometry with the  same comparison and control stars as used by \cite{ah16}.

The observed magnitudes were corrected for the galactic extinction using values 
obtained  from the NED\footnote{\url{https://ned.ipac.caltech.edu}} \citep{sch11} and listed in Table~\ref{tab:optical}.
The magnitudes were converted to flux densities using formula $F=F_0\cdot10^{-mag/2.5}$ with $F_0=4260$ Jy in B,
$F_0=3640$ Jy in V, $F_0=3080$ Jy in R and $F_0=2550$ Jy in I.

The flux densities needed to be corrected for the contribution from the host galaxy of \srcs\ 
at ($z_s=0.944$) and the lens galaxy at $z_l=0.684$.
\cite{jackson00} imaged
this target with the HST through
the $F555W$, $F814W$ and $F160W$ filters and measured the flux density from the face-on spiral galaxy to be  (6$\pm$2), (13$\pm$2) and (15$\pm$2) ($10^{-18}$ erg s$^{-1}$ cm $^{-2}$ \AA$^{-1}$), respectively.
Since \srcs{} 
is classified as a FSRQ \citep{4fgl, pa17}, the host galaxy is likely to be a luminous ($M_K \sim -26.5$) bulge-dominated galaxy \citep[e.g.][]{olguin16}. 
The R-band magnitude of such a galaxy at the redshift of 0.944 would be $\sim 22.5$\,mag.
An early-type galaxy template was taken from \cite{mannu01}, redshifted to $z=0.944$ and integrated over the R-band filter transmission. Then the template was scaled to match the integrated flux density to R = 22.5. The scaled spectrum corresponds to $\sim$ 40\% of the flux densities observed by \cite{jackson00}, i.e. a significant part of the ``spiral galaxy'' surrounding component B could actually be the host galaxy. This is what \cite{fa17} propose based on a high signal-to-noise ratio spectrum of B0218+357. Their spectrum shows gaseous absorption lines at the lens redshift, but no stellar photospheric lines are detected, which led them to propose that the spiral structure belongs to the host galaxy, not the lens. It is impossible to determine the relative contributions of the lens galaxy and the host galaxy from the present data, especially since the latter may also be lensed and absorbed by the former. A simple assumption, that 100\% of the flux densities determined by \cite{jackson00} arise from the lens is used. Thus a fit of a late type (Sa) galaxy template from \cite{mannu01}, redshifted to 0.684, to the \cite{jackson00} flux densities was carried out. Then synthetic photometry was performed through the BVRI bands to the fitted template to obtain flux densities within the aperture for each filter, and these values are reported in Table~\ref{tab:optical}. 
These values were then subtracted from total flux densities.

\begin{table}
    \centering
    \begin{tabular}{ccc}
    \hline
    \hline
    Filter & $A_{X}$ & galaxy flux density [mJy] \\ \hline
    B     &0.25  & 1.4\\
    V     &0.189  & 4.4\\ 
    R     &0.15  & 12\\ 
    I     &0.104  & 31\\
    \hline
    \end{tabular}
    \caption{Galactic absorption values and contribution of the galaxy within the aperture in each filter.}
    \label{tab:optical}
\end{table}

The August 2020 observations with NOT were interrupted by the forest fire on MJD 59084. 
The data from MJD 59083 have lower signal to noise ratio and gradients in the background, resulting in larger than usual reported uncertainties in our analysis. 

\subsection{Radio}

Between January 2017 and January 2019, \srcs{} was frequently observed with KaVA at 22 and 43\,GHz.
A total of 16 sessions were performed during this period. In most cases each session lasted 2 consecutive nights, with a 5--8-hour track at 22\,GHz on the first day and a similar track at 43\,GHz on the following day. 
By default 7 stations (3 from KVN and 4 from VERA) joined each session.
However, occasionally VERA-Mizusawa or VERA-Ishigaki was missing due to local issues. In addition, triggered by the August 2020 campaign, KaVA performed follow-up observations at 43\,GHz for a total of 9 sessions between 2020 August 22 and October 8. The observing time of each follow-up session was 3.5--4 hours and on average 5--6 stations joined. All the data were recorded at 1\,Gbps (a total bandwidth of 256\,MHz with eight 32-MHz subbands) with left-hand circular polarization and correlated by the Daejeon hardware correlator~\citep{lee2015}. The initial data calibration (amplitude, phase, bandpass) was performed using the National Radio Astronomy Observatory (NRAO) Astronomical Image Processing System~\citep[AIPS;][]{greisen2003} based on the standard KaVA/VLBI data reduction procedures~\citep{niinuma2014, hada2017}. Imaging was performed using Difmap software~\citep{shepherd1994} with the standard CLEAN and self-calibration procedures. 

During the 4 KaVA 43\,GHz sessions made on January 15th 2017 (MJD 57768), October 17th 2017 (MJD 58043), November 12th 2017 (MJD 58069) and January 5th 2018 (MJD 58123), additional simultaneous observations of the source with the KVN-only array with a 43\,GHz/86\,GHz dual-frequency recording mode were carried out. A wideband 4\,Gbps mode was used where each frequency band was recorded at 2\,Gbps (a bandwidth of 512\,MHz for each band). 86\,GHz fringes were detected by transferring the solutions derived at 43\,GHz  using the frequency-phase transfer (FPT) technique~\citep[e.g.,][]{algaba2015, zhao2019}. Imaging was carried out in Difmap.

Some of the KaVA/KVN results are presented in \citet{hada2020} together with detailed radio images and analyses at each frequency. See \citet{hada2020} for full details of the KaVA/KVN data reduction and imaging procedures. The typical angular resolution of KaVA (a maximum baseline length $D=$2300\,km) is 1.2\,mas (22\,GHz) and 0.6\,mas (43\,GHz), that of KVN ($D=$560\,km) is 1\,mas at 86\,GHz. Here we report on the whole data set, and investigate the kinematics of the jet.

The OVRO 40-Meter Telescope uses off-axis dual-beam optics and a cryogenic receiver with 2~GHz equivalent noise bandwidth centered at 15~GHz. The double switching technique \citep{1989ApJ...346..566R}, where the observations are conducted in an ON-ON fashion so that one of the beams is always pointed on the source, was used to remove gain fluctuations and atmospheric and ground contributions. 
Until May 2014 a Dicke switch was used to alternate rapidly between  the two beams. Since May 2014 a 180~degree phase switch has been used,  with  a new pseudo-correlation receiver. 
Gain drifts were compensated with a calibration relative to a temperature-stabilized noise diode. The primary flux density calibrator was 3C~286 with an assumed value of 3.44~Jy \citep{1977A&A....61...99B}. DR21 was used as secondary calibrator. \citet{2011ApJS..194...29R} describe the observations and data reductions in detail.
Since the telescope is a single dish, it measures total flux densities integrated over the whole lensed structure: A, B and the Einstein ring.

The 37 GHz observations taken during the August 2020 campaign were made with the 13.7 m diameter Mets\"ahovi radio telescope. 
The observations were ON--ON observations, alternating the source and the sky in each feed horn. 
A typical integration time to obtain one flux density data point was between 1200 and 1400 s. 
The detection limit of the telescope at 37 GHz was on the order of 0.2 Jy under optimal conditions. Data points with a signal-to-noise ratio $< 4$ were treated as non-detections.
The flux density scale was set by observations of DR 21. 
Sources NGC 7027, 3C 274 and 3C 84 were used as secondary calibrators. 
A detailed description of the data reduction and analysis is given in \cite{te98}.
The error estimate in the flux density includes the contribution from the measurement rms and the uncertainty of the absolute calibration. 

\section{Results}  \label{sec:res}

\begin{figure*}
  \centering
  \includegraphics[width=0.9\textwidth]{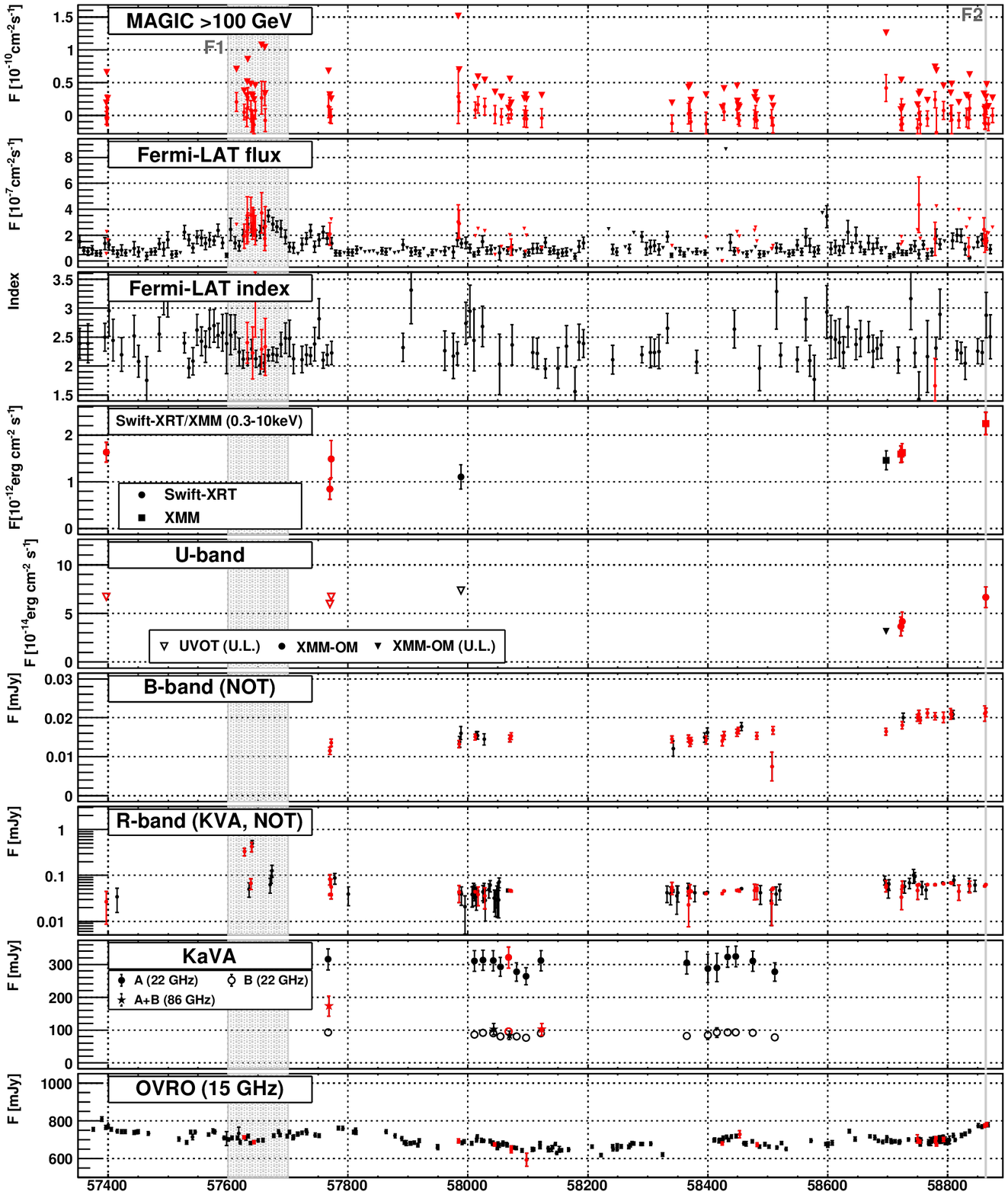}
  \caption{
    MWL light curve of \srcs\ between January 2016 and January 2020.
    From top to bottom:
    MAGIC flux above 100\,GeV, 
    \fermilat\ flux above 0.1\,GeV, 
    \fermilat\ spectral index, 
    X-ray flux in 0.3 -- 10\,keV range (corrected for Galactic absorption) measured with \swift\ and XMM-Newton,
    U-band observations from \swift{}-UVOT and XMM-OM,
    B-band observation from NOT,
    optical observations in R-band with KVA and NOT.
    KaVA VLBI observations at 22\,GHz (filled symbols show A image, empty ones B image) and 86\,GHz (sum of A and B images shown with stars). 
    OVRO monitoring results at 15\,GHz. 
    Flux upper limits are shown  with downward triangles.
    Optical data are corrected for the host/lens galaxy contribution and galactic absorption.
    The points in red are contemporaneous (within 24\,hr slot) with MAGIC observations. 
    The gray filled regions mark the enhanced emission periods F1 and F2.
    \label{fig:mwl_lc}
}
\end{figure*}

 The MWL light curves measured during the monitoring campaign are presented in Fig.~\ref{fig:mwl_lc}. 

\subsection{Search for VHE emission}

\begin{figure*}
  \centering
  \includegraphics[width=0.48\textwidth]{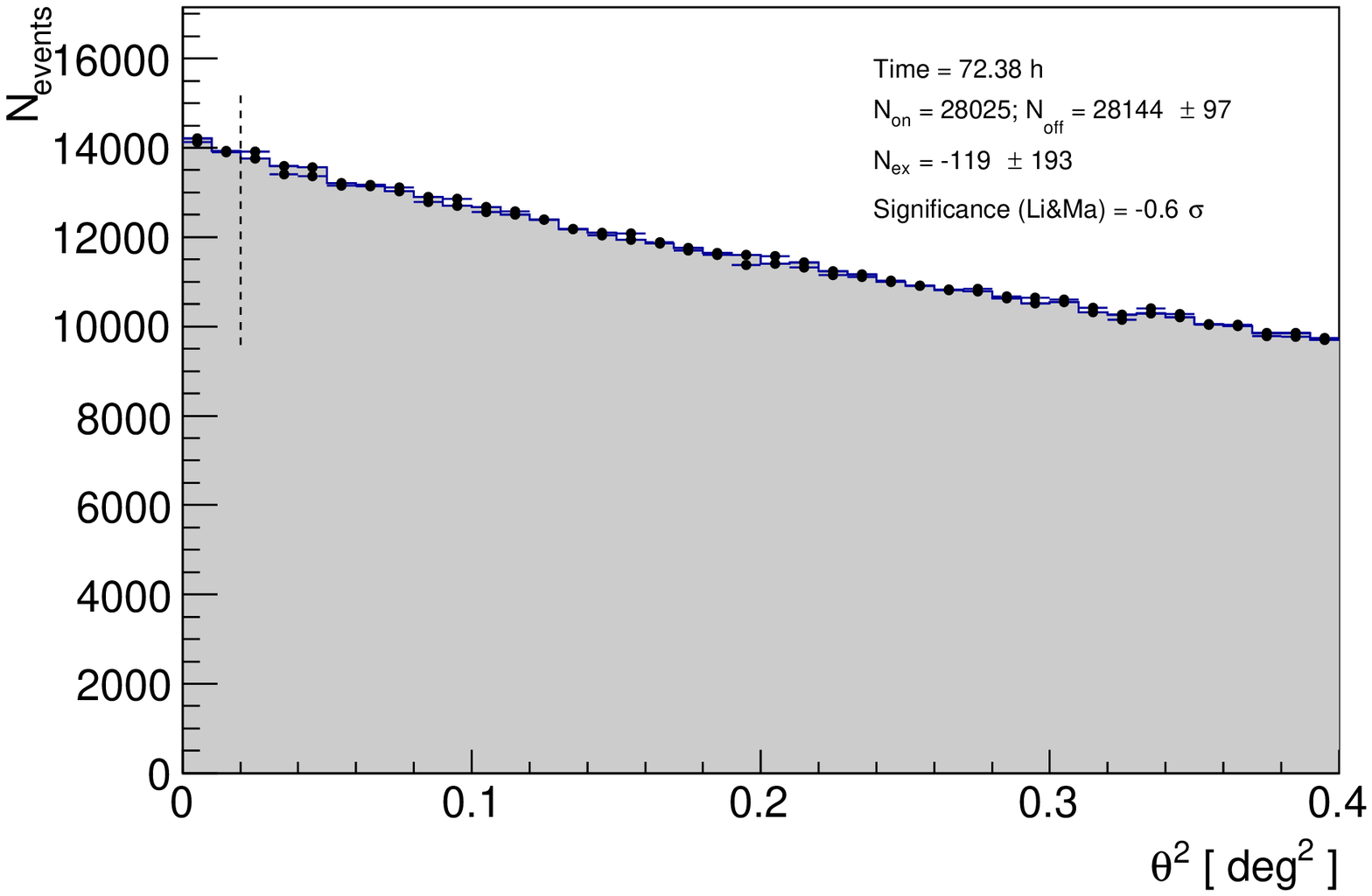}
  \includegraphics[width=0.48\textwidth]{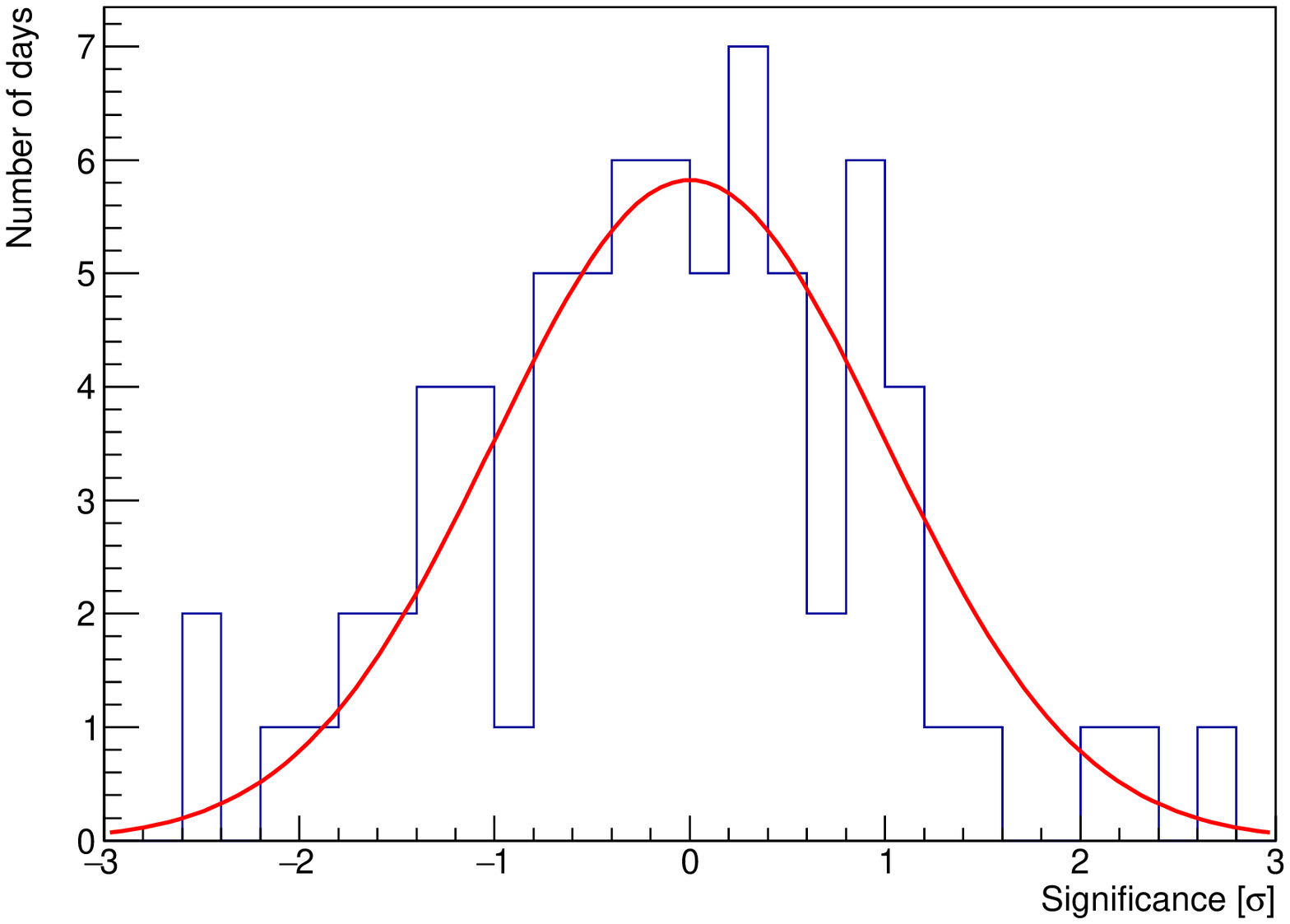}
  \caption{
    Left panel: Distribution of the squared angular distance between the nominal and reconstructed source position of the total dataset of MAGIC data (points) and corresponding background estimation (shaded region). 
    Right panel: distribution of significances of individual nights of MAGIC observations, the red line shows the expected distribution for lack of significant emission, i.e. a Gaussian distribution with a mean of 0 and standard deviation of 1.
}
    \label{fig:magic_sign}
\end{figure*}

No significant VHE gamma-ray emission was found in the total data set of MAGIC monitoring data (see the left panel of Fig.~\ref{fig:magic_sign}).
Due to expected variability of the emission an additional analysis separating the data set into individual nights was performed.
The distribution of the significances of the measured excess is shown in the right panel of Fig.~\ref{fig:magic_sign}, and the upper limits on the flux above 100\,GeV are reported in the top panel of Fig.~\ref{fig:mwl_lc}.
As the source is a known VHE gamma-ray emitter, we also report the nominal flux values on each observation night, however none of them is significant, comparing to the corresponding uncertainty bar. 
The distribution is consistent with the lack of a measurable gamma-ray excess. 
By using the \fermilat{} data,  an additional study was performed to evaluate the expected VHE gamma-ray flux on individual nights (see Appendix~\ref{sec:hardfermi}), however no clear hard GeV states could be identified. 

\begin{figure}
   \centering
   \includegraphics[width=0.96\columnwidth]{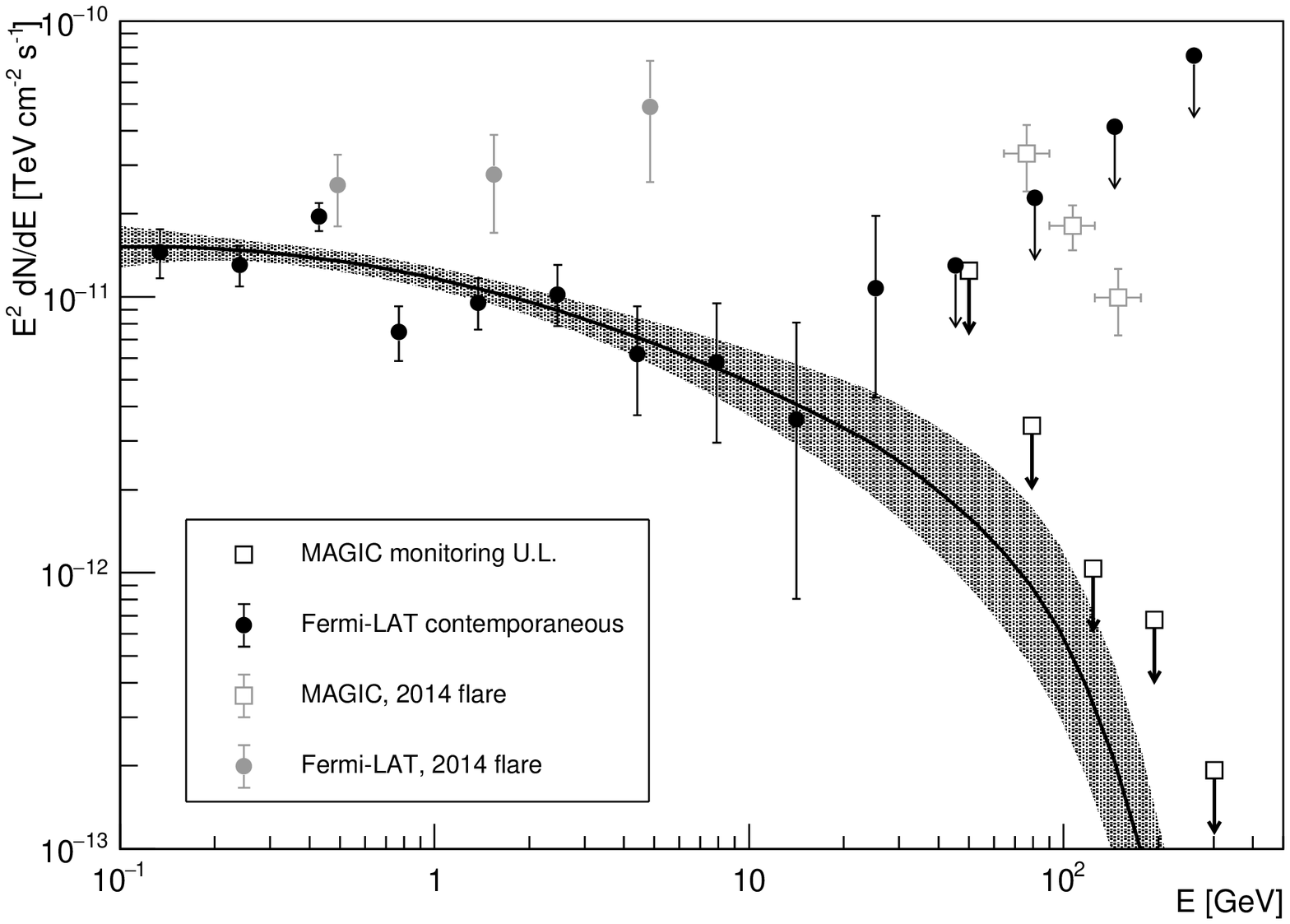}
   \caption{
     Comparison of the MAGIC SED upper limits (black empty squares) with the extrapolation (black line and shaded region) of the \fermilat\ spectrum (black filled circles).
     The extrapolation assumes a power-law behaviour and an extragalactic background light (EBL) absorption following \citet{do11} model.
     For comparison, the \fermilat\ and MAGIC SED during the 2014 flare \citep{ah16} are shown with grey markers.
   }  \label{fig:magic_SED}
\end{figure}

The SED upper limits were computed from the total monitoring sample of the MAGIC observations and compared  with the extrapolation of the \fermilat\ SED (see Fig.~\ref{fig:magic_SED}). 
The \fermilat\ data for this comparison are quasi-simultaneous, i.e. 24\,hr-long time windows  centered on each MAGIC observations are stacked together. 
The MAGIC upper limits are an order of magnitude below the flux observed during the flare in 2014 \citep{ah16}.
However, within the uncertainties of the \fermilat{} extrapolation the upper limits are in agreement with a power-law SED from GeV to sub-TeV range. 

In order to constrain the VHE gamma-ray duty cycle of the source the nights with optimal exposure were selected. 
The data set contains 37 nights with exposure $>100$\,GeV of at least $1.4\times 10^{12}\,\mathrm{cm^{2}s}$ (corresponding to about 1\,hr of observation with a typical effective area of $4\times 10^{8}\,\mathrm{cm^{2}}$). 
All but one of those nights provide 95\% C.L. flux upper limits stronger than the VHE gamma-ray flux observed during the 2014 flare ($5.8\times 10^{-11} \,\mathrm{cm^{-2} s^{-1}}$). 
Following the 2014 event, a duration of an individual flare of at least 2 days was assumed. 
Using Monte Carlo simulations, we related the assumed rate of flares with the corresponding probability of at least one of them being caught in the observation slots of MAGIC. 
We found that the VHE duty cycle is consistent with less than 16 flares per year at 95\% C.L. with a flux $>100$\,GeV of at least $5.8\times 10^{-11} \,\mathrm{cm^{-2} s^{-1}}$. 

\subsection{Enhanced emission periods}
Flux variations were detected across different energy bands during the 4-years-long multi-instrument observations of \srcs\ (see Fig.~\ref{fig:mwl_lc} and Table~\ref{tab:flares}).
\begin{table}
    \centering
    \begin{tabular}{ccc}
    \hline
    \hline
    Tag & MJD & description \\ \hline
    F1     & 57600 -- 57700  & optical and GeV flare \\
    F2     & 58863.7 & X-ray flare\\ \hline
    Aug 2020 & 59071.5 \&  59069.6 & \fermilat{} $>10$\,GeV \\
    \hline
    \end{tabular}
    \caption{List of discussed enhanced emission periods observed during the monitoring (F1 and F2). The campaign in August 2020 was organized after the regular MWL monitoring of the source.}
    \label{tab:flares}
\end{table}
Enhanced GeV emission was observed by \fermilat{} around MJD $\sim$ 57650.
The rise of the GeV emission was gradual. 
In our study, the period from MJD 57600 to MJD 57700 (dubbed as F1) was selected, which covers a time interval where the GeV flux increases and decreases (see Fig~\ref{fig:mwl_lc}).
Based on the obtained light curve, the resulting spectrum reported in this manuscript is not expected to depend strongly on the exact definition of the start and end of this time interval, and that similar results would have been obtained by modifying this time interval by a few days. 
During this time interval, three optical measurements (MJD 57627.2, 57639.2 and 57640.2) yielded a flux nearly an order of magnitude larger than that of the low state of the source. 
Comparing to the 2014 flare discussed in \cite{ah16}, the GeV emission is at a similar level (however with a softer spectrum), but the optical emission is nearly an order of magnitude higher. 
Interestingly the first two points are separated by 12 days, similar to the time delay between the two lensed images of the source. 
The optical flux density between those two measurements returned to the low state level. 
However, due to poor optical sampling of the source, the hypothesis that the two optical flares are indeed the two images of the same flare cannot be validated.
Two of the nights of the enhanced optical activity had simultaneous MAGIC data.
No significant emission was observed (significances of $-0.21\sigma$ and $-0.54\sigma$). 
The flux upper limit above 100\,GeV is $\sim 3 \times 10^{-11}\mathrm{cm^{-2} s^{-1}}$, i.e. constrained to be at least two times below the level observed during the 2014 flare.
Previously, gravitational lensing was used to predict possible sites of gamma-ray flares detected in 2012 and 2014 \citep{2016ApJ...821...58B}. 
They entertained a hypothesis that if both flares were caused by the same plasmoid created in the vicinity of SMBH and traveling toward the radio core then the interaction of the plasmoid with the radio core should be observed around July 2016. 
While OVRO monitoring at 15\,GHz does not show a significant increase in flux density, the predicted interaction coincides with the beginning of F1 in gamma rays, and available observations in R-band show a  significant increase in flux density during the predicted interaction of the plasmoid with the radio core. 
This example illustrates the complexity of studying emissions from these sources but also points to a unique potential and the importance of long-term monitoring of \srcs{} to elucidate the multiwavelength origin of emission.

A hint of enhanced activity was observed in the X-ray band by XMM-Newton on MJD 58863.7 (dubbed as F2). 
The flux density increased by $(44\pm 19)$\% with respect to the previous XMM-Newton measurements.
The contemporaneous MAGIC observations did not yield any significant detection (excess at the significance of $2.1\sigma$).
These observations were used to derive 95\% C.L. upper limit on the flux of $\sim 2.8 \times 10^{-11}\mathrm{cm^{-2} s^{-1}}$ above $100$\,GeV, which is two times below the VHE flux  measured during the flare in 2014.
No significant excess is observed in the MAGIC observations in the two neighbouring nights further suggesting that the marginal excess in MAGIC data during the X-ray flare is a background fluctuation.
No excess of GeV flux was observed during the X-ray flare.
Interestingly, while the optical flux density did not change considerably during F2, a hint of increase in the UV flux density by $(70\pm41)$\% was found. 

\subsection{August 2020 campaign}
Besides the multi-instrument observations described above, \srcs\ is one of the sources that are regularly checked for GeV flares in the \fermilat\ data, and additional observations are organized if flares or hints of flares are found.
On MJD 59071.502 a photon with estimated energy of 59.4\,GeV was observed from the vicinity of the source. 
Additionally, quasi-simultaneous \swift{} observations performed on MJD 59069.566 as well as TJO on MJD 59070.993 showed hints (at $\sim2.5\sigma$ level) of enhanced flux density.
Immediate follow up with the MAGIC telescopes was not feasible, due to the presence of bright moonlight, which would have substantially increased the energy threshold of the observations. 
Instead, similarly to the 2014 event, a multiwavelength campaign was organized at the expected arrival of the B image of the flare, at the assumption that the observed one was the A image. 

\begin{figure}
   \centering
   \includegraphics[width=0.49\textwidth]{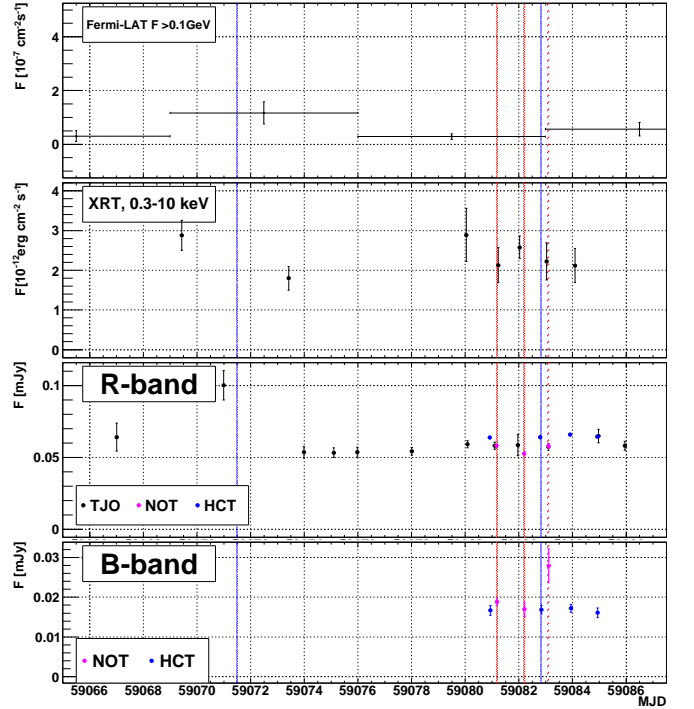}
   \caption{
     MWL light curve of \srcs{} during the August 2020 multiwavelength campaign. Vertical lines: MAGIC observation nights (red) and Fermi-LAT $>$10\,GeV photons(blue). 
   }  \label{fig:aug20_lc}
\end{figure}

The observations are summarized in Fig.~\ref{fig:aug20_lc}. 
A second HE photon with energy of 20\,GeV was observed on MJD 59082.826. 
The association probability of each photon above 10 GeV was assessed with \srcs{} using the standard \fermilat{} tool (\textit{gtsrcprob}).  The whole data set was divided  in the four Point Spread Functions (PSF) classes\footnote{\url{https://fermi.gsfc.nasa.gov/ssc/data/analysis/documentation/Cicerone/Cicerone_Data/LAT_DP.html}}. This is needed since there are relevant PSFs differences between the four classes and those differences have an important effect on the association probability. 
During the August observations, only these two photons above 10 GeV were detected with an association probability higher than 80\%. The probability of the two photons being associated to the source is 99.91\% and 86.88\% for the first and the second photon respectively. 
The time difference of these two photons is 11.324 days, which is curiously consistent with the previously measured delay between the two images. 

In order to evaluate statistical chance probability of occurrence of HE photons close in time with the emission of the source in a broader time scale is investigated. 
120 such photons spanning the total observations of the source by \fermilat{} (MJD=54683 -- 59299) are obtained.
Conservatively assuming the time window for the second photon of $\pm1$\,day (motivated by the spread of radio delays of $\sim10-12$\,days) a chance probability of 5.2\% is obtained. 
The time delay of 11.324\,days is also within the $1\sigma$ uncertainty of 
that measured from 2012 \fermilat{} high state ($11.46\pm0.16$\,days, \citealp{ch14}). 
For such a narrow window the corresponding chance probability is $0.83\%$.

In the optical range a hint of increase of the R-band emission ($2.5\sigma$ difference to the previous point and $4.2\sigma$ to the next) occurred close to the arrival time of the first \fermilat\ photon. 
The observations during the planned monitoring at MJD=59081 -- 59085 were performed with higher cadence with additional instruments (NOT and HCT). 
The NOT measurement at MJD$=59083.1$ is: $2.5\sigma$ above the previous HCT observation, $2.4\sigma$ above the following HCT measurement. 
Similarly, comparing the enhanced NOT point to the previous NOT measurement at MJD$=59082.2$, the difference shows a similar weak hint of $2.4\sigma$. 
%
The B-band flux density increase was not accompanied by a similar R-band increase at the expected time of arrival of the second component of the flare. 
This would favour the interpretation of the B-band increase as a statistical fluctuation. 

While some small hint of variability ($2.3\sigma$ difference) can be seen between the first two \swift\ points, no variability can be seen during the expected time of arrival of the delayed component. 
This is understandable since the delayed component is expected to have a lower flux density at the peak, hence any small variability present in the leading emission can be easily missed in the trailing one. 

\begin{figure}
   \centering
   \includegraphics[width=0.96\columnwidth]{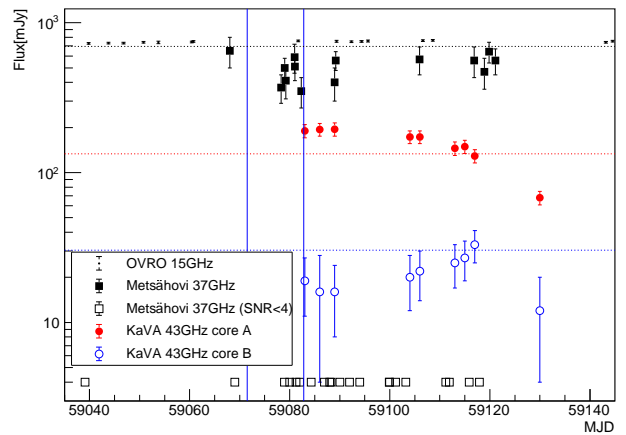}
   \caption{
     Radio follow up light curve of the August 2020 campaign of \srcs{} with OVRO (black points) and KaVA (red filled and blue empty circles for core image A and B respectively). 
     Black squares show the significant 37\,GHz Mets\"ahovi flux densities (empty squares shows the times of observations that resulted in signal-to-noise ratio below 4). 
     The dotted lines show the average flux density from the monitoring period between January 2017 and December 2018.
     The vertical blue lines show the times of arrival of the two HE \fermilat{} photons during the August 2020 campaign. 
     }\label{fig:aug20kava}
\end{figure}

In Fig.~\ref{fig:aug20kava} radio follow-up light curves of the August 2020 campaign obtained with OVRO at 15\,GHz and KaVA at 43\,GHz is shown. 
For KaVA data in which A and B are spatially separated, the radio flux density for the core (the brightest component) was measured  in each of A and B images. 
The core of A in August 2020 is found to be significantly brighter than the average flux density level in 2017--2018, and subsequently it shows continuous decrease in flux density at least until the end of our follow-up period (October 8), where the 43\,GHz core flux density reaches a lowest level since the start of our KaVA monitoring from 2017. 
In contrast, we caution that the observed light curve of the weak component B is less defined because the observing conditions of KaVA follow-up sessions in 2020 were generally severe compared to the regular sessions in 2017-2018 (shorter integration time and smaller number of stations). 
In Fig.~\ref{fig:aug20kava} there may be a significant amount of missing flux density in the light curve for B. This prevents us from cross-correlating the light curves of A and B. 
The 15\,GHz OVRO monitoring did not cover the period in which the 43\,GHz flux density decreased. 
The OVRO data 3 weeks before and 2 weeks after the KaVA flux density minimum show consistent flux densities. 
Mets\"ahovi data cover the period in which the flux density measured by KaVA starts to decay, however higher uncertainties and a larger integration region prevents sensitive probing of variability. 
%
%

\subsection{Radio jet image}\label{sec:radiojet}

\begin{figure*}
   \centering
   \includegraphics[width=0.95\textwidth]{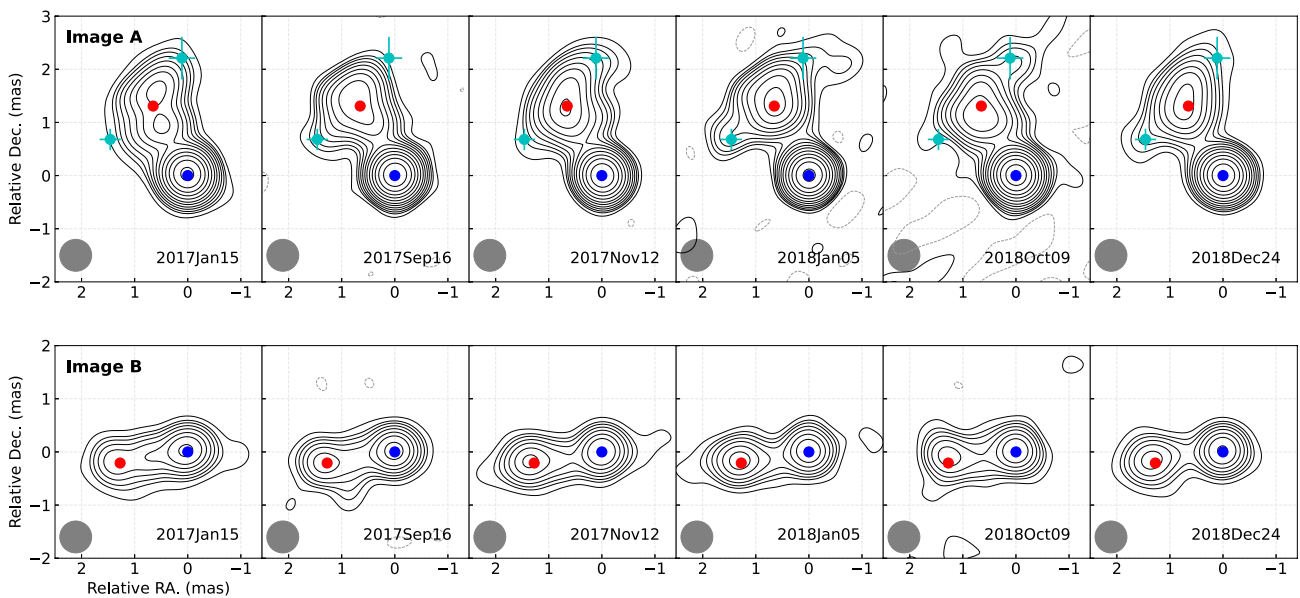}
   \caption{
     Selected KaVA images at 43\,GHz spanning 2 year period. Top panel and bottom panels are images A and B of the source. For all images, contours start from $-1$ (dotted lines), 1, 2, ... times 1.8\,mJy\,beam$^{-1}$ (approximately 3$\sigma$) and increase by factors of $2^{1/2}$.
     Blue and red dots represent the average (over all the images) position of the core and jet components. Two cyan points in image A represent the average positions of the sideway components, which were obtained based on 5 epochs where the sideway extension was clearly detected. 
     For all the points, the position uncertainties were estimated based on the scatter from multiple epochs. Gray circle represents the smoothing kernel of the image. 
   }  \label{fig:kava_images}
\end{figure*}

In Fig.~\ref{fig:kava_images} the evolution of the radio images of the source between January 2017 and December 2018 is presented. Here  KaVA 43\,GHz images are shown since the highest angular resolution is available. 
In addition to the core, a strong jet component is clearly detected (at $\sim$1.5/1.3\,mas from the core in A/B respectively, which corresponds to the projected distance between the two components of the order of 10\,pc) in all the images. No additional knots have been observed throughout the observations. The jet direction is different in Image A and Image B, however this is a geometrical effect caused by the lensing.

In order to examine the kinematics of the jet component, the core and jet in the KaVA 43\,GHz images were fitted by two 2D elliptical Gaussians, and the core-jet separation was measured as  function of time. As can be seen in Fig.~\ref{fig:kava_images}, the core-jet angular separations for both A and B images are quite stable over $\sim$2 years of our observing period. Assuming that the same jet component is traced over 2 years, a simple linear fit to the data is performed. Best-fit values of $0.06\pm0.03$\,mas/yr and $0.04\pm0.04$\,mas/yr are obtained for Jet-A and Jet-B, respectively (corresponding to $3.1\pm1.5c$ and $2\pm2c$). 

In addition to the bright core and jet, the KaVA images of A also reveal diffuse extension in the direction perpendicular to the jet (see also \citealp{bi03,hada2020}).  Two Gaussian models were additionally fitted to each KaVA 43\,GHz image of A to characterize the positions of these extended structures. Since the sideways structures are generally diffuse and weak,  reasonable fitting results were obtained on these features only for 5 epochs  (MJD=58069, 58123, 58434, 58450, 58476; 2017 November 12, 2018 January 5, November 12, November 28 and December 24) where the image quality was relatively high and SNR$\sim$4--10 were obtained for the fitted sideways components. In Fig.~\ref{fig:kava_images}, the positions of the these additional features (averaged over the 5 epochs) are plotted in cyan. With respect to the core, the apparent `opening angle' of this perpendicular extension (the angle between two vectors from the red point to each cyan point in Fig.~\ref{fig:kava_images}) is estimated to be $\sim$62$^{\circ}$ (if only the extension of the main jet is considered, the opening angle is roughly a half of this). Possible proper motions in these components were also searched for, but both of the components are essentially stationary, similar to the main jet component. 


Regarding the mas-scale jet morphology during the August 2020 campaign, higher noise levels in the KaVA images (due to shorter integration time, smaller number of stations, higher humidity in the summer season) than in the 2017--2018 sessions made our image analysis more challenging (especially for the image B). Nevertheless, the overall radio morphology was quite similar to that in 2017--2018 (Fig.~\ref{fig:kava_images}). While the core of A in August 2020 was significantly brighter than the average flux density level in 2017--2018 (see Fig.~\ref{fig:aug20kava}), no clear ejection of new components from the core during our KaVA follow-up period was found.

\subsection{Lens geometry model} \label{sec:lensing}

The observations of \srcs{} with the Hubble Space Telescope (HST) show that the lensing galaxy is isolated pointing to a simple gravitational lens potential. 
The mass distribution of the lens has been shown to be well represented by a Singular Isothermal Sphere (SIS) model \citep{2004MNRAS.349...14W,2011AstL...37..233L,2005MNRAS.357..124Y,2016ApJ...821...58B}. 
The SIS model of \srcs{} predicts a time delay of $\sim$ 10 days and a magnification ratio of $\sim$ 3.6. The predicted magnification ratio  fits the observed ratio between radio images. 
However, the time delay is $\sim$1 day shorter as compared to the time delay measured at gamma rays \citep{ch14,2016ApJ...821...58B}.  

\citet{hada2020} used a Singular Elliptical Power-law (SEP) model and parameters fixed to the values obtained by \citet{2004MNRAS.349...14W}. The SEP model provides a higher rate of change in the lens potential. As a result, the model with the same image positions predicts longer time delays and a larger magnification ratio between the images. The SEP model adapted by \citet{hada2020} predicts a time delay of 11.6 days for the core, which matches better the observed time delay at gamma rays. However, the predicted magnification ratio is $\sim \,$4.8, which significantly deviates from a reported average value of 3.5 from a broad range of radio observations \citep{1995MNRAS.274L...5P}. 

At first parameters of the lens model presented in \cite{2016ApJ...821...58B} are adopted, which reconstructed the positions of the lensed images with an accuracy of $1\,$mas. 
The previous model was based on 15~GHz radio observations \cite{1995MNRAS.274L...5P}. 
Here, the reconstruction of the lens model is further improved by using high resolution KaVA observations of the radio core and jet listed in Table~\ref{tab:radio_positions}.

\begin{table*}
\begin{center}
\begin{tabular}{ |c|c|c|c|c|c|c|c|c| } 
 \hline
 \hline
 Component &Image& $RA$                 & $DEC$                 & $\Delta_{MODEL}$  & Source        & Time Delay    & $\mu$     & Ratio \\ 
           &    & [mas]                 & [mas]                 & [mas]             & [mas]         & [days]        &           &       \\
\hline
 Core      & A  & 0 $\pm$ 0             & 0   $\pm$ 0           & 0.067             & (90.0,37.1)   & 10.36         & 2.72      & 3.81  \\ 
           & B  & 309.144$\pm$0.015     & 127.450$\pm$0.029     &  0                &               &               & -0.71     &       \\ 
\hline
    Jet    & A  & 0.681$\pm$0.031       & 1.331$\pm$0.045       & 0.036             &(89.06,36.21)  & 10.30         & 2.74      & 3.67 \\ 
           & B  & 310.444$\pm$0.014     & 127.253$\pm$0.038     & 0.260             &               &               & -0.75     &       \\ 

 \hline
\end{tabular}
\caption{
    Positions of radio images observed by KaVA and the lens model predictions. 
    The positions of the images are referenced to the lensed image A (0,0). 
    Position errors are estimated based on the scatter of fitted positions 
    based on the assumption that all of the components are stationary.
    The positions ($RA,DEC$) are shown for the observed lensed images A and B for the core and jet components. 
    The image A was modeled by 4 Gaussians (core, main jet, left wing, right wing).  
    The lensed image B was modeled by 2 Gaussians (core, jet). 
    Table reports averaged positions over 5 epochs (2017Nov12, 2018Jan05, 2018Nov12, 2018Nov28, 2018Dec24). 
    The $\Delta_{MODEL}$ represents a difference between the predicted and observed positions of the lensed images,
    as well as reconstructed positions of the core and jet in the source plane in respect to the lens center. 
    Table also shows time delays, magnifications, and magnification rations predicted using the best SIS lens model.
  }
\label{tab:radio_positions}
\end{center}
\end{table*}

A softened power-law potential \citep{2001astro.ph..2341K,2016ApJ...821...58B} is used, which includes an Einstein radius, a scale radius of a flat core ( $s$ in mas), a projected axis ratio ($q$), and a power-law exponent ($\alpha$). Similarly, like in \cite{2016ApJ...821...58B}, the best lens model of softened power-law potential resulted in $s\sim 0$, $q \sim 1$, and $\alpha \sim 1$. Thus, the model reduced to the SIS potential.

\begin{figure}
   \centering
   \includegraphics[width=0.96\columnwidth]{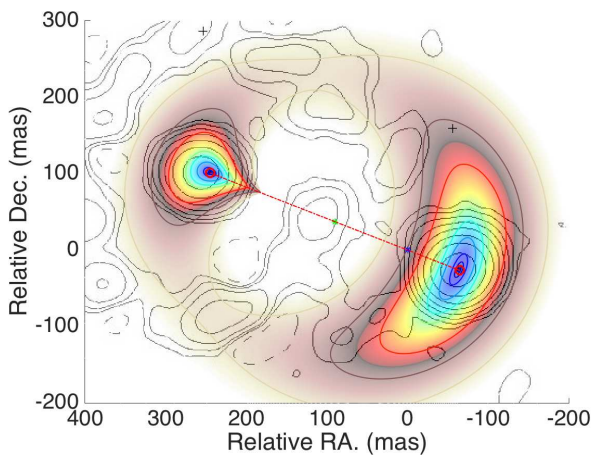}
   \caption{
     Compilation of observations and lens model predictions.
     The color map shows the Fermat potential of the lens model for the best reconstructed position of the radio core.
     The images form at the extreme of the Fermat surface (blue). 
     The gray contours show the lensed images of the core and the Einstein ring observed at 1.687\,GHz \protect\cite{2004MNRAS.349...14W}. 
     The image is centered at the reconstructed position of the lens indicated as the blue asterisk. 
     The red open circles correspond to the reconstructed position of the images of the jet(A on the right, and B on the left). 
     The red dashed-dotted line connects the images. 
     For the SIS lens model, the source (green asterisk) is located at half distance between images.  
     The observed and reconstructed images of the radio core are also shown, however they are superimposed with the red open circles. 
   }\label{fig:lens_model}
\end{figure}

Figure~\ref{fig:lens_model} shows the radio observations with the prediction of the SIS model including the Fermat potential, predicted positions of the lensed images, as well as reconstructed position of the the jet and core. 
The image is centered at the position of the lens located at (244.55,100.85) with respect to image A.

\begin{figure}
  \centering
  \includegraphics[trim=60 10 80 50, clip, width=0.23\textwidth]{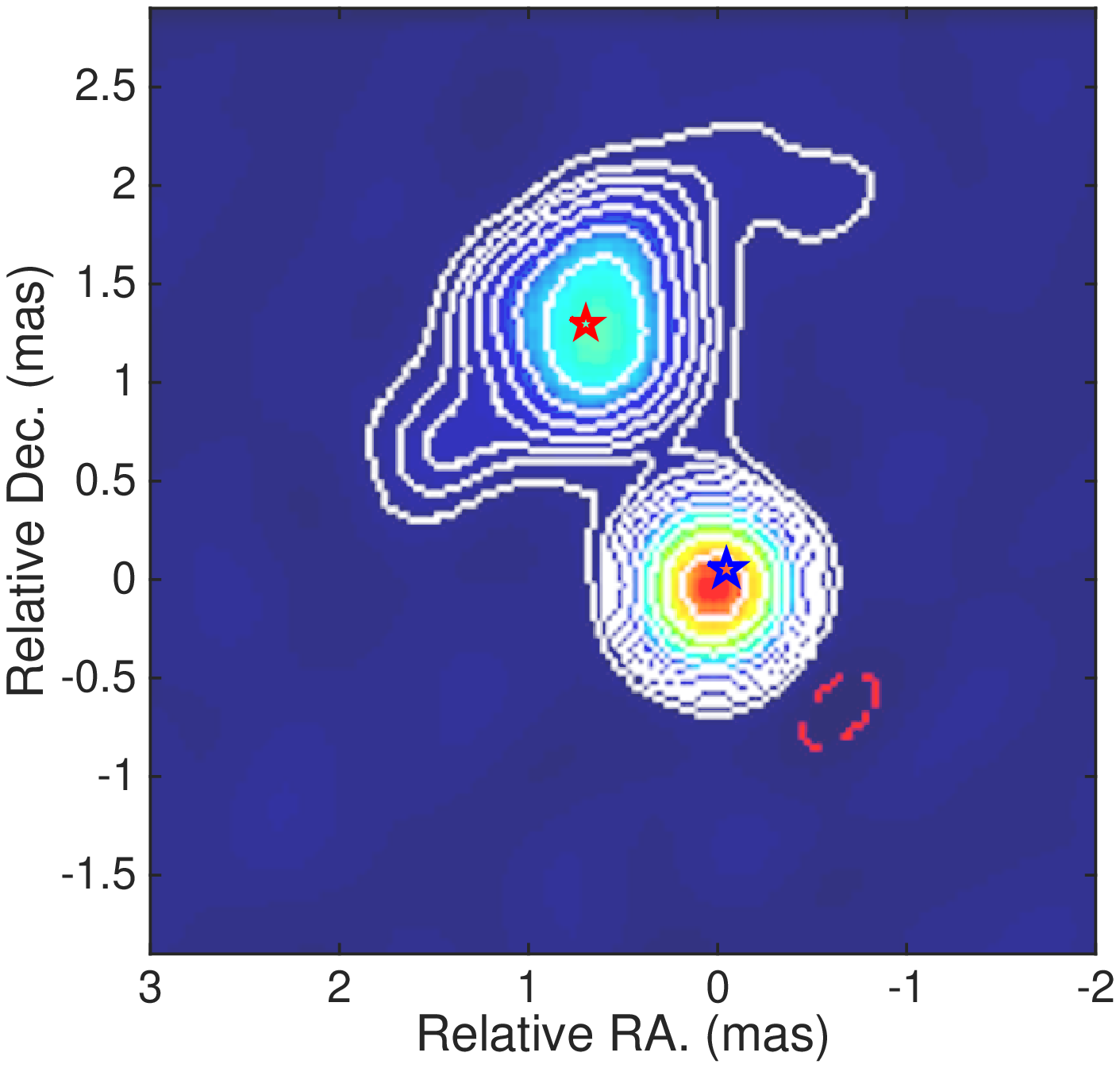}
  \includegraphics[trim=60 10 80 50, clip, width=0.23\textwidth]{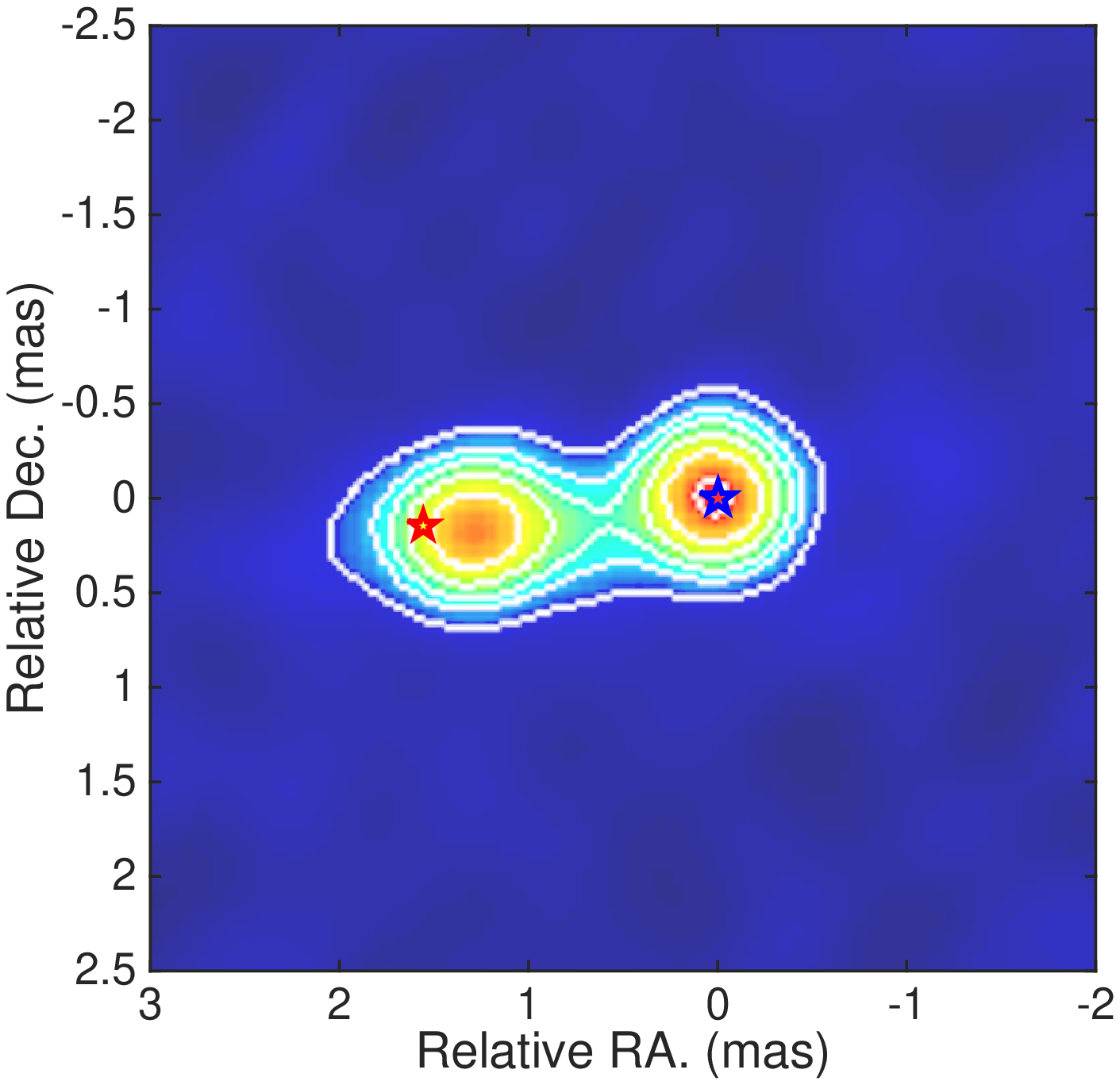}
  \caption{
  Comparison of the KaVA 43\,GHz image taken on MJD=54470 (2018 January 5),  A in left panel, B in right panel compared with the reconstructed positions of the lens model images (stars)
}
    \label{fig:lens_images}
\end{figure}

Only positions of the radio images were used to find the best fit, 
as the observed time delays and magnification ratios are a subject of discussion. 
Figure~\ref{fig:lens_images} shows the position of reconstructed images as red and green open circles.
The model reconstructs the positions of the lensed images with average accuracy of 0.03$\,$mas for the radio core and 0.15$\,$mas for the jet. 

The projected distance between reconstructed positions of the core and jet is $1.2\pm 0.1\,$mas, which correspond to $9.8\,$pc in the source plane, consistent with the result obtained by \citet{hada2020}. 
The SIS model predicts a time delay of 10.36 days for the core and 10.30 days for the jet. 
The shorter time delay for the jet indicates that the radio jet is positioned toward the lens center as argued by \cite{2016ApJ...821...58B} based on observations of the Einstein radius at 1.687 GHz \cite{2004MNRAS.349...14W}. The Einstein ring forms from emission of the kpc-scale jet aligned with the lens center. The Einstein ring observed at 1.687 GHz \citep{2004MNRAS.349...14W} is also added to Figure~\ref{fig:lens_model}.  

The predicted time delays reported here are calculated using $H_0 = 67.3\pm \mathrm{1.2\,km\,s^{-1}\,Mpc^{-1}}$ \citep{ssdc_pccs1}. 
Note that the time delay is inversely proportional to $H_0$. 
Thus, larger values of $H_0=73.3^{+1.7}_{-1.8}\mathrm{\,km\,s^{-1}\,Mpc^{-1}}$ as reported by \citet{2020MNRAS.498.1420W} would result in an even shorter time delay, and as such would further increase the discrepancy between the lens models and observations.
The Hubble parameter estimation based on gravitationally-induced time delays of variable sources with relativistic jets is in particular prone to biases as the emission can originate from multiple sites, which can introduce systematical errors \citep{2015ApJ...799...48B}.

Both the SEP \citep{hada2020} and improved SIS reconstruction based on KaVA observations, provide accurate reconstruction of the positions of the lensed images 
and consistent distance separation of the radio core and jet components.   
However, the predictions of these two models differ in terms of time delay and magnification ratios. 
The SIS scenario predicts magnification ratios of 3.81 and 3.67 for the core and jet, respectively. 
The flux densities of the lensed images at 43 GHz reported in Table~3 \citep{hada2020} indicate magnification ratios of 3.7 and 3.2 for the core and jet, respectively. 
For comparison, the SEP model predicts magnification ratios of 5 and 4.84 for the core and jet, respectively. 

In principle, magnification ratio could be affected by microlensing or ``substructure lensing'' such as compact clumps in the lens galaxy \citep{sb16}.
To match the prediction of the magnification ratio of $\sim5$ of the SEP model with the observed $\sim 3.5$, either the brighter image A would have to be magnified by a factor of 1.35, or image B would have to be de-magnified by the same factor.    
Moreover, the lensed images of both jet and core would have to be magnified/de-magnified simultaneously by a similar factor. 
As a result, the diameter of the perturbing mass needs to be $\gg 10\,$pc, precluding microlensing. 
In principle, possible clumps of tens of pc of a giant molecular cloud (GMC) (see e.g. Chevance et al. 2020) could result in additional moderate amplifications by a factor of 1.5 \citep[see][]{sb16}.

Interestingly, the observed magnification ratio fits the prediction of the SIS model well. 
The SIS model predicts correctly not only the values but also the degree to which the magnification ratio of the core is greater than the magnification ratio of the jet.  
However, the time delay of 10.3 days predicted by the SIS model is one day shorter that the time delay measured at gamma rays.  
Such a discrepancy in the time delay could be explained if there is an offset of $\sim50\,pc$ between sites of radio and gamma-ray emission \citep[for a review see][]{2018PhR...778....1B}.  

The degeneracy between the SIS and SEP lens models could be broken by precise measurement of the radio time delay. The time delay obtained in the SIS model depends only on $H_0$ and the image angular separation. Thus, the SIS model can be ruled out if the radio time delay is not 10.3 days. 

However, measuring time delay at radio with an accuracy of hours is difficult as radio observation of B2 0218+35 shows almost no variability, in addition to gaps between the observations. 
Further high resolution KaVA observations combined with detailed modeling of the lens could elucidate the true potential of the lens and test the clump scenario. 
KaVA observations provide well-resolved images of both the jet and core, thus providing multiple tests of the lens potential.   
One of the predictions of the SIS model is that the diameter of the Einstein ring is equal to the distance between the lensed images of the source. 
The current KaVA observations show that the distance between the lensed images of both the core and jet are equally within the range of uncertainty of the observations - as such, consistent with the SIS model. More precise observations, or detailed predictions of the SEP model on the expected difference between the lensed images of the core and jet could help exclude the SEP model. 
Moreover, elliptical lens models should result in formation of the odd number of images \citep{1994JMP....35.5507G,2007MNRAS.377.1623Z,2010GReGr..42.2011P}. 
Thus, detection of the third image in the vicinity of the predicted lens center could provide further constraints on the model of the lens. 

Here, we focused on the two most general lens models, namely SIS and SEP, and on the possibility to break the degeneracy between them. However, a potentially more general class of models might be required to reconstruct both the time delay and the magnification ratio.

The in-depth observations of the object B2 0218+35, combined with a precise model of the lens, have the potential to provide unique insights on the origin and site of the gamma-ray emission, the Hubble constant, or substructures in the lensing galaxy.

\subsection{Modelling of dust in the lens}

A Galactic absorption of N$_{H}$=5.56$\times$10$^{20}$ cm$^{-2}$ was adopted from the Leiden/Argentine/Bonn (LAB) survey \citep{Kalberla05}.
The X-ray flux density, corrected for the Galaxy absorption, in the (0.3--10) keV band is $f$=(1.53$\pm$0.11)$\times$10$^{-12}$ erg cm$^{-2}$ s$^{-1}$ for the low flux density state, and $f$=(2.25$\pm$0.24)$\times$10$^{-12}$ erg cm$^{-2}$ s$^{-1}$ for the high-flux density state from XMM-Newton observations. 

In order to evaluate and correct the effect of additional absorption in the host or lens galaxies an approach similar to \citet{ah16} is applied. The higher sensitivity of XMM-Newton compared to the \swift\ telescope allows us to study a few alternative models of absorption and intrinsic source spectrum and select between them. For the investigations of the low state three cases are considered: (a) no additional absorption, (b) absorption at the host, (c) absorption at the lens. In the case (b) the absorption will affect the total emission observed from the source (in both images). In the case (c), since the two images cross different parts of the lens galaxy, the absorption would be different for them. There are reasons to believe that in such a situation the absorption would mainly affect the brighter, A, image (see the discussion in \citealp{ah16}). Therefore in the case (c) the observed emission is assumed to be composed of two ``virtual'' sources located at the nominal location of \srcs{} in the sky. The first component  is affected only by Galactic absorption, and the second one is additionally absorbed by a hydrogen column density  (N$_{H,z}$)  at the redshift of the lens ($z$=0.68). 

Two spectral models are investigated: a simple power law and a log parabola (defined as $F(E)=k(E/E_0)^{-\Gamma}$ and  $F(E)=k(E/E_0)^{-(\alpha+\beta \log(E/E_0))}$, where $E_0$ = 1 keV in all the models).
For the absorption,  the  Tuebingen-Boulder ISM absorption model \citep{Wilms00} available in XSPEC was adopted.  
The column density  N$_{H,z}$  and the  parameters  of the log-parabola or power-law components were fitted by imposing that  the values of $\alpha$, $\beta$, or  $\Gamma$ are the same for the two components, and that the normalization of the component with only Galactic absorption is a factor   $0.71/2.72=0.261$  
(corresponding to magnification ratio, see Section~\ref{sec:lensing}) lower than the normalization of the component with also internal absorption. 

\begin{figure}
   \centering
   \includegraphics[width=0.95\columnwidth, trim=10 5 50 150, clip]{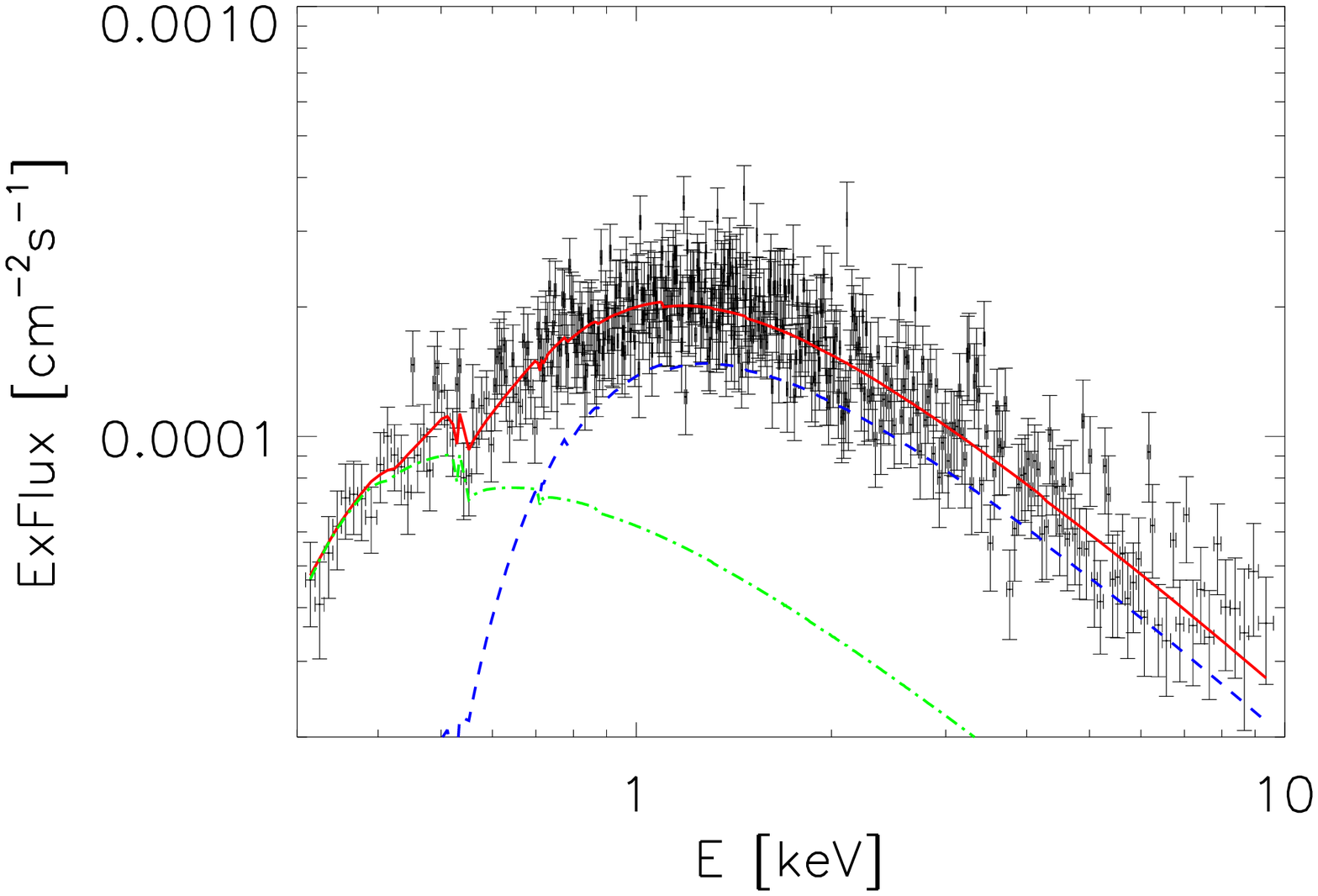}
   \includegraphics[width=0.95\columnwidth, trim=10 5 50 150, clip]{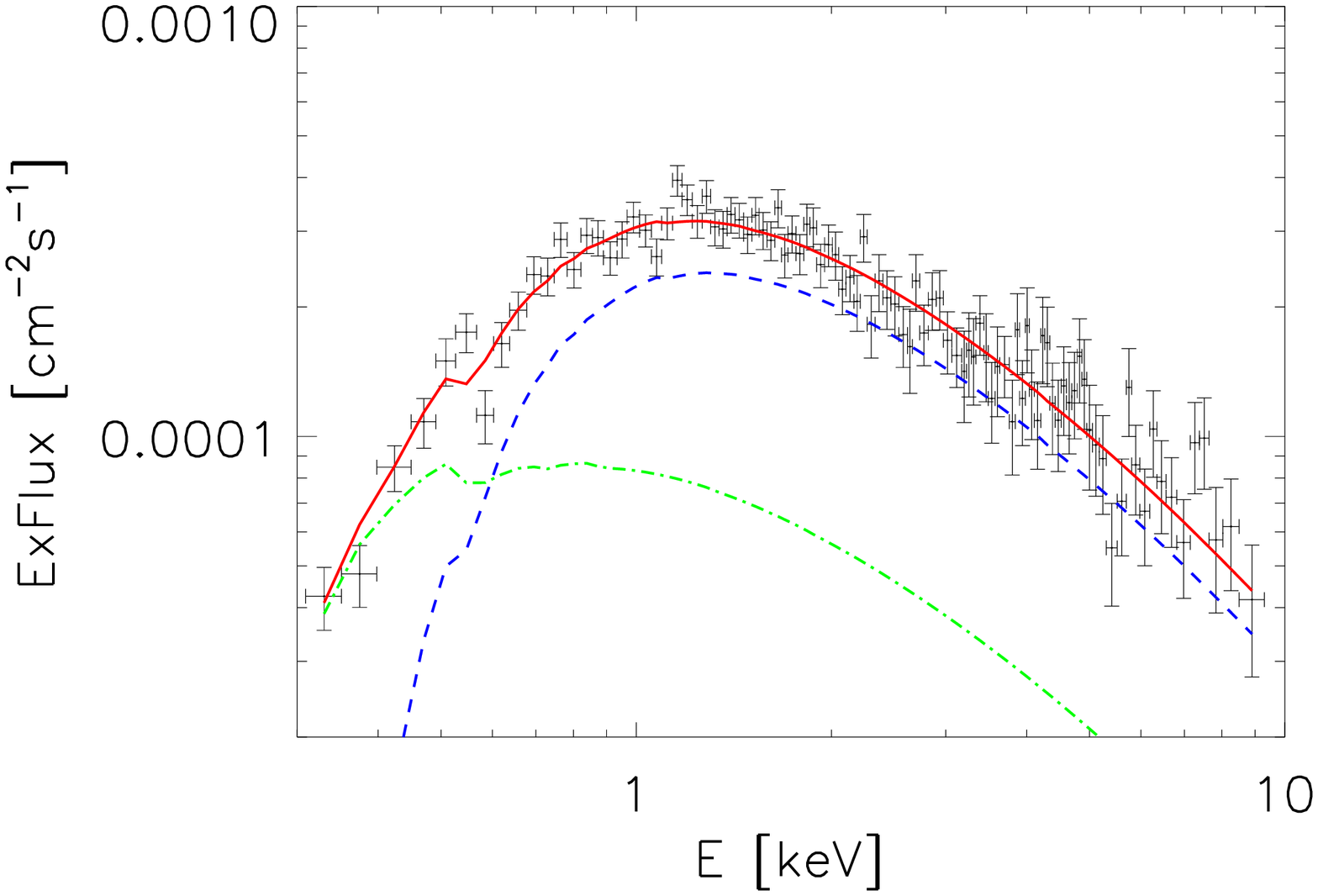}
   \caption{ Differential energy flux of \srcs{} folded with the response of XMM-Newton observed from low-state pointings (top panel) and from the X-ray flare (bottom panel).
   Points show the observed rate, while lines show the LP model for image A (blue dashed), image B (green dot-dashed) and total emission (red solid).
   }  \label{fig:xmm_spectra}
\end{figure}

The results of the fits  are summarized in Table \ref{XMM_fit}. 
\begin{table*}
\small
\begingroup
\setlength{\tabcolsep}{1.5pt} 
\renewcommand{\arraystretch}{1} 
\begin{threeparttable}
  \caption{Best-fit parameters of the XMM-Newton and \swift\ analysis. }
  \label{XMM_fit}
  \begin{center}
    \begin{tabular}{cccccccccccc}
      \hline
    \hline
obs. ID & Exp. time & Model &$\alpha$ & $\beta$  & $\Gamma$&$k_1$ &$k_2$&$z$  & N$_{H,z}$   & $\chi^2$/N$_{dof}$  \\

(1) &     (2) & (3)  & (4)       & (5)  & (6) &(7) & (8)  & (9)  &  (10) &(11)\\  
 \hline
%
%
 \textbf{low state} & \textbf{27.6} & \textbf{LP} &\textbf{1.96$\pm$0.05} & \textbf{0.17$\pm$0.05} &\textbf{--} &\textbf{0.71}& \textbf{2.73$\pm$0.11}& \textbf{0.68}&\textbf{8.10$\pm$0.93}  & \textbf{375.0/373} \\ 
 low state & 27.6 & PL &-- &-- &2.06$\pm$0.03   & 0.72 &   2.78$\pm$0.10&0.68 & 8.83$\pm$0.82& 386.6/374\\
 low state & 27.6 & LP & 1.37$\pm$0.08 & 0.72$\pm$0.10 & --& 2.43$\pm$0.09& --  &0.94& 1.88$\pm$0.49& 398.9/373\\
 low state & 27.6 & LP & 1.05$\pm$0.03 & 1.09$\pm$0.05 & --& 2.14$\pm$0.03& --  &--& --& 413.4/374\\
 low state & 27.6 & PL &--&--&1.95$\pm$0.03 & 2.98$\pm$0.07&--&0.94& 5.16$\pm$0.27& 442.8/374 \\
 \hline
\textbf{0850400601} &\textbf{10.3}&\textbf{LP} &\textbf{1.58$\pm$0.16} & \textbf{0.53$\pm$0.15} &\textbf{--} &\textbf{0.96}& \textbf{3.68$\pm$0.36} & \textbf{0.68}&\textbf{5.4$\pm$1.8} & \textbf{90.5/94} \\
0850400601 &10.3&LP+low&1.89$\pm$0.40* & 0.17$\pm$ 0.42* &-- & -- & 2.14$\pm$0.42* & 0.68& 7.0$\pm$2.3* & 89.7/94 \\
\hline
\swift\ (2016-2017) & 9.4 & PL &-- &-- &  1.62$\pm$0.13 & 0.47 & 1.79$\pm$0.18 & 0.68 & 8.10 & 7.7/9\\

\swift\ (2020) & 20.4 & PL &-- &-- &  1.83$\pm$0.06 & 0.80 & 3.07$\pm$0.15 & 0.68 & 8.10 & 33.2/34\\
\hline
\end{tabular}
\begin{tablenotes}
      \small
      \item \textbf{Notes.} Columns: 
      (1) observation identifier (``low state'' corresponds to combined 0850400301, 0850400401 and  0850400501 epochs of XMM-Newton observations);  
      (2) exposure time  filtered for good time intervals in ks; 
      (3) log-parabola (LP)/power-law (PL) spectral model; 
      (4) and (5) LP spectral index and curvature; 
      (6) PL spectral index; 
      (7) and (8) normalization of the LP/PL model at 1 keV in units of 10$^{-4}$ cm$^{-2}$s$^{-1}$keV$^{-1}$ of B and A image respectively; 
      (9) redshift of the absorber; 
      (10) column density of the absorber in units of 10$^{21}$ cm$^{-2}$; 
      (11) reduced $\chi^2$
      Final selected model for each dataset is marked with bold face. 
      * - only the additional flaring component
    \end{tablenotes}
  \end{center}
  \end{threeparttable}
\endgroup
\end{table*}
The simple log-parabola model, without an additional absorption is not sufficient to explain the XMM-Newton data ($\chi^2/N_{dof}=413.4/374$). 
Including an additional absorption at the lens for the brighter image the $\chi^2$ improves by $38.4$ at the cost of one additional degree of freedom (hydrogen column density at the lensed image A). 
The model is compared with observed XMM-Newton rates in Fig.~\ref{fig:xmm_spectra}. 

Since not all the investigated models are nested, to compare them Akaike Information Criterion \citep{ak74} is used. 
For the case of $\chi^2$ statistics the relative difference of AIC parameter of two models can be computed as:
$\Delta \mathrm{AIC}=2\Delta n_p +\Delta \chi^2$. 
The relative likelihood of the models can be computed (see e.g. \citealp{bah11}) as: 
$p=\exp(\Delta \mathrm{AIC}/2)$. 
Including the absorption at $z=0.68$ the intrinsic curvature of the emission is preferred. 
The difference of $\chi^2$ values is $11.6$ for one additional parameter (describing the curvature) corresponds to $p=0.008$ relative likelihood of the power law model to the log parabola model. 
On the other hand, comparing models with absorption at $z=0.68$ and $z=0.94$, with the same number of free parameters in both models, the former has a lower $\chi^2$ value by 23.9.  
Therefore the model with absorption at the source is only $1.8\times10^{-5}$ as likely as the model with the absorption at the lens. 
Summarizing, for the low state of the source, the preferred model of the emission involves an intrinsic log-parabola spectrum and absorption by a column density of $(8.10\pm 0.93_{\rm stat}) \times 10^{21}\mathrm{cm^{-2}}$ at the lens. 

Comparing to the result obtained in \citet{ah16} $24\pm 5_{\rm stat} \times 10^{21}\mathrm{cm^{-2}}$, the absorption obtained in this work is more precise, but also lower, and both values are consistent in the broad range derived by \citet{mr96} ($5-50 \times 10^{21}\mathrm{cm^{-2}}$). 
The actual absorption in the lens might have changed if the region emitting X-rays has moved along the jet, or changed its size compared to the observations in 2014. 
However it is equally likely that additional systematics (assumption of the intrinsic spectral model, flux density and spectral variability, absorption in the other image of the lens and in the host galaxy) affected one or the other measurement. 

The proper correction for the absorption and lensing during the MJD 58863.7 high X-ray state is more uncertain. 
Since the observations are separated by 140 days from the previous X-ray flux measurement, is not clear if the X-ray high state was a short duration flare, or a longer time scale high state.
In a case of a short flare the observations might have happened when the corresponding image A or image B reached the observer, resulting in a different absorption. 
On the other hand, if the enhanced state was significantly longer than the $\sim11$ days delay between the two images the observed emission should be the average from both images, similar to the case of the low state. 
This is further supported by the fact that the data collected by \swift\ for MJD 59069 -- 59108 show also higher X-ray fluxes, similar to the last XMM-Newton measurement.  
To analyze MJD 58863.7 data of XMM-Newton the assumption that the observed increase in the X-ray emission was over a longer time scale is applied, therefore the log-parabola spectral model was used with absorption of the brighter image and fixed flux ratio between both components. 
Such a model describes sufficiently well the observations ($\chi^2/N_{dof}=90.5/94$), and provides a somewhat harder and more curved X-ray spectrum than during the low state. 
The derived $N_{H,z}$ is roughly consistent (at $1.3\sigma$ level) with the values obtained from the low state fit. 
An alternative model has been tested in which the high state emission is a sum of the low state emission, with the spectral shape and flux fixed to the low-state-fitted values and an additional flaring component with an absorption at $z=0.68$ (i.e. the flare originating from the brighter image A). 
The fit results for such an additional flaring component are reported in ''LP+low'' row of Table~\ref{XMM_fit}. 
The two investigated models for the flaring state are not nested, however they have the same number of free parameters and result in nearly the same $\chi^2$, therefore neither can be rejected. 
For the sake of simplicity the same absorption model is used for the flaring state as for the low state. 

As discussed in Section \ref{sec:swift-data}, two combined spectra are used for the spectral study using \swift{} data during 2016-2017 and August 2020. Due to the restriction aroused by \swift\ sensitivity, only two spectral models within the scenario of case (c) are investigated. In addition to the assumption implemented for XMM-Newton observations, N$_{H,z}$ is also assumed to be equal to $8.10\times10^{21}$\,cm$^{-2}$. The log-parabola model cannot describe the observed spectrum better than the power law model. The results are presented in Table~\ref{XMM_fit}.



\section{low state broadband SED modelling} \label{sec:model}

\begin{figure}
   \centering
   \includegraphics[width=0.96\columnwidth]{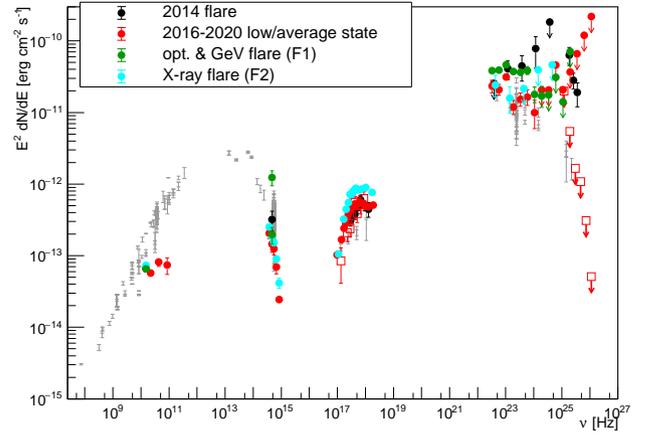}
   \caption{ Multiwavelength SED of \srcs\ in different periods: average state of the monitoring (excluding optical flare data) in red (for clarity \swift{} and MAGIC data are shown with empty symbols, while XMM-Newton and \fermilat{} and the rest of MWL data are shown with full symbols), X-ray flare in cyan, and optical/GeV flare in green. 
   For the case of optical flare and minimum and maximum optical flux density during investigated period are plotted. 
   For comparison historical data (obtained from SSDC service) are plotted in gray, while the 2014 flare \citep{ah16} is in black. 
   }  \label{fig:mwl_seds}
\end{figure}

The multiwavelength behaviour of the source in different states is summarized in Fig.~\ref{fig:mwl_seds}.
During the low state the GeV emission was significantly lower than during the 2014 flare described in \citet{ah16}, and even during the F1 enhanced state the GeV spectrum was softer than in 2014. 
Except for the strong optical flares in F1, the rest of the MWL SED does not vary strongly with respect to the 2014 flare and historical measurements.
The difference in reported radio measurements to the historical ones is most likely caused by much smaller integration region (only the inner jet of the source) achieved in radio interferometry with KaVA. 

For the average state a detailed modelling is performed. 
The emission is modeled in the framework of an External Compton (EC) model, which is a common scenario for FSRQs. 
There is growing evidence that the main target for FSRQs EC process is the reprocessed Dust Torus (DT, see e.g. \citealp{co18,be19}). 
Moreover, even while no VHE gamma rays has been detected from the \srcs\ during the monitoring period, the source is a known emitter in this band \citep{ah16}, again supporting EC-DT scenario.

The recent measurement of the accretion disk luminosity of $L_d=4.3\times 10^{44}\mathrm{erg\,s^{-1}}$  \citep{2021ApJS..253...46P} is used.
The value is close to the $L_d=6\times 10^{44}\mathrm{erg\,s^{-1}}$ estimated by \citet{gh10} and applied in \citet{ah16}. 
Using the updated value of $L_d$ the sizes of the BLR and DT are computed using the scaling laws of \citet{2009MNRAS.397..985G}, resulting in 
$R_{\rm BLR}=6.6\times 10^{16}$\,cm, and $R_{\rm DT}=1.6\times10^{18}$\,cm. 
The temperature of the DT is set to 1000\,K and its luminosity to $0.6 L_d$. 
A conical jet geometry is used, with half-opening angle of $1/\Gamma$, where $\Gamma$ is the Lorentz factor of the jet. 

For the modeling the Doppler factor of the jet $D=\Gamma=15$ is assumed.
The electron energy distribution (EED) is assumed to follow a power law with an index of $p_1$ up to $\gamma_b$, where $\gamma_b$ is the Lorentz factor of the electrons for which the time scale for the dominating energy loss process is equal to the dynamic scale (see \citealp{acc21} for details).
Above such a cooling break the EED steepens by 1 up to $\gamma_{\max}$, which is determined from balancing the acceleration gain with the dominating energy loss process. 
The radiation processes are calculated using the \texttt{agnpy}\footnote{\url{https://github.com/cosimoNigro/agnpy/}} code \citep{agnpy}, which implements the synchrotron and Compton processes following the prescriptions described in \citet{dm09,fi16}.
While the $\gamma_b$ and $\gamma_{\max}$ are calculated assuming Thompson regime of the inverse Compton scattering, the actual spectra are computed using the full Klein-Nishina cross-section formula. 

\begin{figure*}
   \centering
   \includegraphics[width=0.7\textwidth]{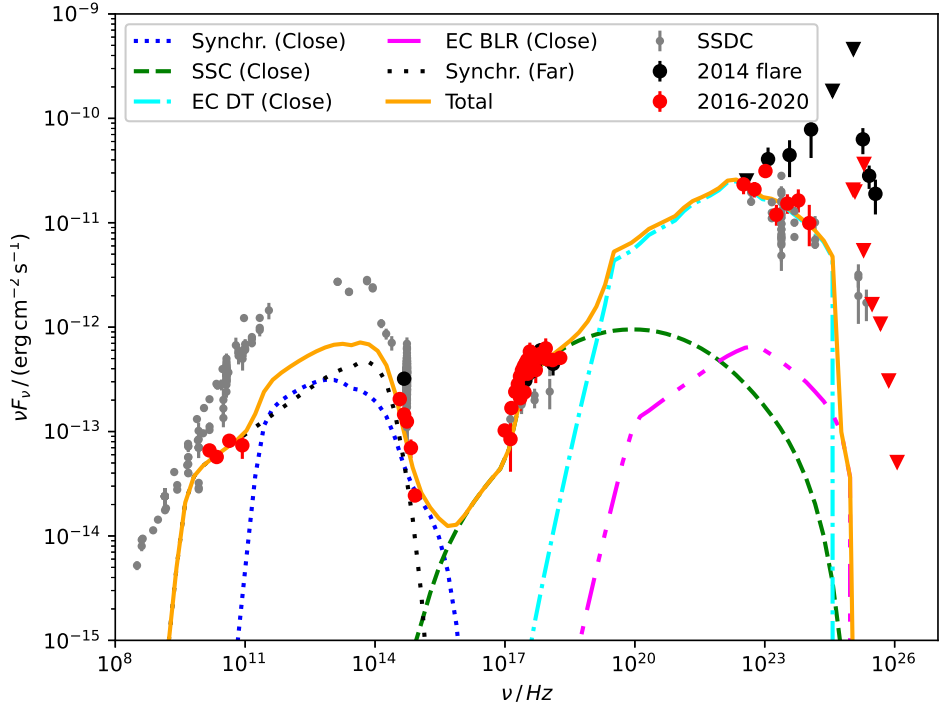}
   \caption{ Multiwavelength SED of \srcs\ composed of contemporaneous data to the MAGIC observations (red points), historical data from SSDC (gray) and data from the 2014 flare (\citealp{ah16}, in black) compared with the broad-band model derived from a two-zone SSC+EC scenario (with parameters reported in Table~\ref{tab:table_model}). 
   Optical, UV and X-ray data are corrected for the Galactic absorption, optical data are in addition corrected for the host/lens galaxy contribution. 
   The lensing magnification, absorption in the lens galaxy and EBL attenuation are corrected for in the model curves.
   For the closer region, dotted curve is the synchrotron emission, dashed the SSC, dot-dashed EC on DT, dot-dot-dashed EC on BLR. 
   For the farther region, long-dotted is the synchrotron emission. 
   The total emission is shown with an orange line. 
   }  \label{fig:mwl_model}
\end{figure*}

The emission region (hereafter ``Close'' region) responsible for the high energy bump is assumed to be located at the distance of $d_1=2\times 10^{17}$\,cm, i.e. a factor of a few more distant than the size of the BLR, but deep in the DT radiation field. 
The model is confronted with the observations in Fig.~\ref{fig:mwl_model}, taking into account the magnification induced by the lensing (using the strong lensing magnifications derived in Section~\ref{sec:lensing}), and the absorption of emission from one of the images in the lens. 
The possible effect of microlensing is not corrected for, however we expect it to have a minor influence on the long-term average spectrum. 
The free and derived parameters are summarized in Table~\ref{tab:table_model}. 

\begin{table*}
    \centering
    \begin{tabular}{c|c|c|c|c|c|c|c||c||c|c|c|c}
    Region &  $\delta$ & $d$ [cm] & $\xi$ & $B$ [G] & $u_e\mathrm{[erg\,cm^{-3}]}$ & $p_1$ & $\gamma_\mathrm{min}$ &|& $p_2$ &  $\gamma_\mathrm{break}$ & $\gamma_\mathrm{max}$ & $r_b$ [cm] \\ \hline
    Close & 15 & $2\times10^{17}$& $5\times10^{-7}$ & $0.11$ & $0.7$ & $2.4$ & 50 & |& $3.4$ & 1500 & 26 000 & $1.3\times10^{16}$ \\ 
    Far & 15 & $3\times10^{20}$& $6\times10^{-10}$ & $3.2\times10^{-3}$ & $4\times 10^{-7}$ & $2.4$ & 2 &|& - & - & 51 000 & $2\times10^{19}$
        \end{tabular}
    \caption{Parameters used for the modelling: Doppler factor $\delta$, distance of the emission region $d$, acceleration efficiency $\xi$, magnetic field $B$, electron energy density $u_e$, EED: slope before the break: $p_1$, minimum Lorentz factor $\gamma_\mathrm{min}$, slope after the break $p_2$,  the Lorentz factor of the break $\gamma_\mathrm{break}$, maximum Lorentz factor $\gamma_\mathrm{max}$, co-moving size of the emission region $r_b$.
    Free parameters of the model and derived parameters are put on the left and right side of the vertical line respectively.
    For the case of ``Far'' region $B$ and $u_e$ are tied with equipartition condition. }
    \label{tab:table_model}
\end{table*}
The gamma-ray emission is explained as EC process on DT photons (which is also the dominating energy loss process of the electrons). 
On the other hand, according to the model, the X-ray emission is mostly caused by SSC process. 
The synchrotron emission corresponding to the ``Close'' region can (largely) explain the optical and the rapidly falling UV emission. 

However, the region is too compact for explaining the low-frequency radio emission which is heavily absorbed in the ``Close'' region  by synchrotron-self-absorption. 
Such low-energy emission is expected to originate from a larger scale jet.
A commonly applied solution is the assumption of two emission regions (see e.g. \citealp{acc20} and references therein). 
Therefore, motivated also by the radio knot observed by KaVA, a second region (hereafter ``Far'') is added, located at the distance of 100\,pc. 
The distance of the emission region is motivated by (deprojected) distance of the jet component in the KaVA image. 
The low-energy slope in this region is set to $2$, and  equipartition (i.e. $u_e = B^2/(8\pi)$) is applied.
Then the magnetic field strength and the acceleration coefficient are fixed to the values explaining the broadband synchrotron emission. 
The two emission regions are assumed not to be co-spatial (``Far'' region is more distant in the jet) and thus contrary to e.g. \citet{acc20} are not interacting with each other.   
The ``Far'' region is distant and large-enough such that the dominating energy loss process is the synchrotron cooling, which, again due to size of the region and low values of the acceleration efficiency, does not introduce a cooling break up to the maximum reached energies. 
Inverse Compton emission of the ``Far'' region is negligible in comparison to the ``Close'' region (the region is beyond the DT radiation field for EC process to play any role, and the energy density of the electrons is too low for SSC to be effective). 

Combination of both emission regions can describe remarkably well the whole broadband emission of the source. 
In fact, the previous modelling of the source, presented in \citet{ah16}, also suggested two-zone model for this source. 
That modeling was however used to explain the flaring episode of the source, and neglected the radio and microwave emission from the large-scale jet. 
It is therefore likely that the three regions contribute to the time-variable, broad-band emission of the source: large scale jet responsible for the radio and microwave emission, emission region within DT responsible for the broadband, high-energy, low state emission of the source and a third region (or a sub-region of the second one) in which VHE and HE gamma ray flares occur. 
As the low state modelling attributes most of the radio emission to the ``Far'' region, it is curious to observe radio variability in August 2020 campaign KaVA data over time scales of tens of days (see Fig.~\ref{fig:aug20_lc}). 
Such variability might not be connected with the source itself, but rather with the  absorption and milli-lensing effects of large scale structures in the lensing galaxy.
In Appendix~\ref{sec:flares} a possible scenario is discussed for explaining the MWL flares seen from the source during the monitoring. 

\section{Conclusions} \label{sec:conc}

Broadband (radio, optical-UV, X-ray, gamma ray) monitoring of \srcs{} has been performed. 
The monitoring was aimed at the detection of a VHE gamma-ray flare of the source in time periods selected to allow additional follow up at the expected time of arrival of the second image. 
The deep exposure of 72\,hrs of data did not reveal low-state VHE gamma-ray emission and constrained it to be less than about of an order magnitude below the level observed during 2014 flare. 
VLBI radio images obtained with KaVA show clear core-jet structure in both lensed images. 
No significant movement of the VLBI radio features was seen. 
No significant variability has been seen in KaVA images during the 2016-2019 monitoring, however the follow up of August 2020 campaign showed a clear decay of core flux density in image A.
The radio data have been used to improve the lens modelling to evaluate image magnifications and time delays for the core and jet component of the source.
Precise measurements of the X-ray spectrum with XMM-Newton instrument was used to evaluate the absorption in the lens, and fit the hydrogen column density of the lens in the image A to a value of $(8.10\pm0.93) \times 10^{21}\mathrm{cm^{-2}}$. 

The low-state broadband emission of the source can be described with a two-zone model, in which the electron energy distribution shape is determined self-consistently from the cooling, acceleration and dynamical time scales. 
Most of the radio and FIR emission is explained to originate from a large region with size/location motivated by the radio jet component. 
UV (and partially optical) data are explained as the synchrotron emission of the smaller region, that is also responsible for the gamma-ray emission (produced in EC scenario) and X-ray emission (generated via SSC process).

No short VHE gamma-ray flares have been observed in the night-by-night analysis. 
Comparison with the \fermilat\ state of the source shows that it is unlikely that the source has reached a comparable flare to the one of 2014 during the monitoring. 
The MAGIC data were used to place a 95\% C.L. limit on the VHE gamma-ray duty cycle of the source: below 16 flares per year.  

Monitoring data have revealed however a few flares/hints of enhanced states in optical, X-ray and gamma-rays, during which no VHE gamma-ray emission was detected. 
While the limited MWL data and variability during enhanced periods do not allow us to properly model the enhanced states, a plausible scenario explaining qualitatively the change of behaviour of the source during those states by change of the basic parameters of the model is presented. 

Additional MWL campaign triggered by hints of enhanced emission in gamma-rays, X-rays and optical has been also discussed. 
Unfortunately lack of MAGIC data on the predicted night of the flare prevents us from drawing a firm conclusion on possible hardening of the electron energy distribution during the campaign. 

While the primary goal of the MWL monitoring of the source has not been achieved due to in general low gamma-ray activity of the source in the last years, the campaign resulted in multiple interesting results, and observations of a few interesting events. 
Since the achieved constraints on the low-state VHE gamma-ray emission approach the extrapolation of the GeV emission, it is expected that the future Cherenkov Telescope Array \citep{2013APh....43....3A} will allow us to study it in detail.

\section*{Acknowledgements}
We would like to thank the Instituto de Astrof\'{\i}sica de Canarias for the excellent working conditions at the Observatorio del Roque de los Muchachos in La Palma. The financial support of the German BMBF, MPG and HGF; the Italian INFN and INAF; the Swiss National Fund SNF; the ERDF under the Spanish Ministerio de Ciencia e Innovaci\'{o}n (MICINN) (FPA2017-87859-P, FPA2017-85668-P, FPA2017-82729-C6-5-R, FPA2017-90566-REDC, PID2019-104114RB-C31, PID2019-104114RB-C32, PID2019-105510GB-C31,PID2019-107847RB-C41, PID2019-107847RB-C42, PID2019-107988GB-C22); the Indian Department of Atomic Energy; the Japanese ICRR, the University of Tokyo, JSPS, and MEXT; the Bulgarian Ministry of Education and Science, National RI Roadmap Project DO1-268/16.12.2019 and the Academy of Finland grant nr. 320045 is gratefully acknowledged. This work was also supported by the Spanish Centro de Excelencia ``Severo Ochoa'' SEV-2016-0588 and CEX2019-000920-S, and ``Mar\'{\i}a de Maeztu'' CEX2019-000918-M, the Unidad de Excelencia ``Mar\'{\i}a de Maeztu'' MDM-2015-0509-18-2 and the ``la Caixa'' Foundation (fellowship LCF/BQ/PI18/11630012) and by the CERCA program of the Generalitat de Catalunya; by the Croatian Science Foundation (HrZZ) Project IP-2016-06-9782 and the University of Rijeka Project uniri-prirod-18-48; by the DFG Collaborative Research Centers SFB823/C4 and SFB876/C3; the Polish National Research Centre grant UMO-2016/22/M/ST9/00382; and by the Brazilian MCTIC, CNPq and FAPERJ.
The \textit{Fermi} LAT Collaboration acknowledges generous ongoing support
from a number of agencies and institutes that have supported both the
development and the operation of the LAT as well as scientific data analysis.
These include the National Aeronautics and Space Administration and the
Department of Energy in the United States, the Commissariat \`a l'Energie Atomique
and the Centre National de la Recherche Scientifique / Institut National de Physique
Nucl\'eaire et de Physique des Particules in France, the Agenzia Spaziale Italiana
and the Istituto Nazionale di Fisica Nucleare in Italy, the Ministry of Education,
Culture, Sports, Science and Technology (MEXT), High Energy Accelerator Research
Organization (KEK) and Japan Aerospace Exploration Agency (JAXA) in Japan, and
the K.~A.~Wallenberg Foundation, the Swedish Research Council and the
Swedish National Space Board in Sweden.
Additional support for science analysis during the operations phase is gratefully
acknowledged from the Istituto Nazionale di Astrofisica in Italy and the Centre
National d'\'Etudes Spatiales in France. This work performed in part under DOE
Contract DE-AC02-76SF00515.
We thank Director of Indian Institute of Astrophysics for allotting us observing
time with HCT under DDT. 
We also thank the staff of IAO, Hanle and CREST, Hosakote, that made HCT observations  possible. 
The facilities at IAO and CREST are operated by the Indian Institute of Astrophysics, Bangalore.
The Joan Oró Telescope (TJO) of the Montsec Astronomical Observatory (OAdM) is owned by the Catalan Government and operated by the Institute for Space Studies of Catalonia (IEEC). 
This research has made use of data from the OVRO 40-m monitoring program which was supported in part by NASA grants NNX08AW31G, NNX11A043G and NNX14AQ89G, and NSF grants AST-0808050 and AST-1109911, and private funding from Caltech and the MPIfR.
S.K. acknowledges support from the European Research Council (ERC) under the European Unions Horizon 2020 research and innovation programme under grant agreement No.~771282.
This research has made use of the NASA/IPAC Extragalactic Database (NED), which is funded by the National Aeronautics and Space Administration and operated by the California Institute of Technology.
This publication makes use of data obtained at the Mets\"ahovi Radio Observatory, operated by Aalto University in Finland.
Part of this work is based on archival data, software or online services provided by the Space Science Data Center - ASI.
We would like to thank the anonymous journal reviewer for his/her comments that helped to improve the manuscript. 

\section*{Data Availability}
The data used in this article were accessed from the MAGIC telescope \url{http://vobs.magic.pic.es/fits/}. 
The derived data generated in this research will be shared on reasonable request to the corresponding author.


\section*{Affiliations}
$^{1}$ {Instituto de Astrof\'isica de Canarias and Dpto. de  Astrof\'isica, Universidad de La Laguna, 38200, La Laguna, Tenerife, Spain} \\
$^{2}$ {Universit\`a di Udine and INFN Trieste, I-33100 Udine, Italy} \\
$^{3}$ {National Institute for Astrophysics (INAF), I-00136 Rome, Italy} \\
$^{4}$ {ETH Z\"urich, CH-8093 Z\"urich, Switzerland} \\
$^{5}$ {Institut de F\'isica d'Altes Energies (IFAE), The Barcelona Institute of Science and Technology (BIST), E-08193 Bellaterra (Barcelona), Spain} \\
$^{6}$ {Japanese MAGIC Group: Institute for Cosmic Ray Research (ICRR), The University of Tokyo, Kashiwa, 277-8582 Chiba, Japan} \\
$^{7}$ {Technische Universit\"at Dortmund, D-44221 Dortmund, Germany} \\
$^{8}$ {Croatian MAGIC Group: University of Zagreb, Faculty of Electrical Engineering and Computing (FER), 10000 Zagreb, Croatia} \\
$^{9}$ {IPARCOS Institute and EMFTEL Department, Universidad Complutense de Madrid, E-28040 Madrid, Spain} \\
$^{10}$ {Centro Brasileiro de Pesquisas F\'isicas (CBPF), 22290-180 URCA, Rio de Janeiro (RJ), Brazil} \\
$^{11}$ {Universit\`a di Padova and INFN, I-35131 Padova, Italy} \\
$^{12}$ {University of Lodz, Faculty of Physics and Applied Informatics, Department of Astrophysics, 90-236 Lodz, Poland} \\
$^{13}$ {Universit\`a di Siena and INFN Pisa, I-53100 Siena, Italy} \\
$^{14}$ {Deutsches Elektronen-Synchrotron (DESY), D-15738 Zeuthen, Germany} \\
$^{15}$ {INFN MAGIC Group: INFN Sezione di Torino and Universit\`a degli Studi di Torino, 10125 Torino, Italy} \\
$^{16}$ {Max-Planck-Institut f\"ur Physik, D-80805 M\"unchen, Germany} \\
$^{17}$ {Universit\`a di Pisa and INFN Pisa, I-56126 Pisa, Italy} \\
$^{18}$ {Universitat de Barcelona, ICCUB, IEEC-UB, E-08028 Barcelona, Spain} \\
$^{19}$ {Armenian MAGIC Group: A. Alikhanyan National Science Laboratory} \\
$^{20}$ {Centro de Investigaciones Energ\'eticas, Medioambientales y Tecnol\'ogicas, E-28040 Madrid, Spain} \\
$^{21}$ {INFN MAGIC Group: INFN Sezione di Bari and Dipartimento Interateneo di Fisica dell'Universit\`a e del Politecnico di Bari, 70125 Bari, Italy} \\
$^{22}$ {Croatian MAGIC Group: University of Rijeka, Department of Physics, 51000 Rijeka, Croatia} \\
$^{23}$ {Universit\"at W\"urzburg, D-97074 W\"urzburg, Germany} \\
$^{24}$ {Finnish MAGIC Group: Finnish Centre for Astronomy with ESO, University of Turku, FI-20014 Turku, Finland} \\
$^{25}$ {Departament de F\'isica, and CERES-IEEC, Universitat Aut\`onoma de Barcelona, E-08193 Bellaterra, Spain} \\
$^{26}$ {Armenian MAGIC Group: ICRANet-Armenia at NAS RA} \\
$^{27}$ {Croatian MAGIC Group: University of Split, Faculty of Electrical Engineering, Mechanical Engineering and Naval Architecture (FESB), 21000 Split, Croatia} \\
$^{28}$ {Croatian MAGIC Group: Josip Juraj Strossmayer University of Osijek, Department of Physics, 31000 Osijek, Croatia} \\
$^{29}$ {Japanese MAGIC Group: RIKEN, Wako, Saitama 351-0198, Japan} \\
$^{30}$ {Japanese MAGIC Group: Department of Physics, Kyoto University, 606-8502 Kyoto, Japan} \\
$^{31}$ {Japanese MAGIC Group: Department of Physics, Tokai University, Hiratsuka, 259-1292 Kanagawa, Japan} \\
$^{32}$ {Saha Institute of Nuclear Physics, HBNI, 1/AF Bidhannagar, Salt Lake, Sector-1, Kolkata 700064, India} \\
$^{33}$ {Inst. for Nucl. Research and Nucl. Energy, Bulgarian Academy of Sciences, BG-1784 Sofia, Bulgaria} \\
$^{34}$ {Finnish MAGIC Group: Astronomy Research Unit, University of Oulu, FI-90014 Oulu, Finland} \\
$^{35}$ {Croatian MAGIC Group: Ru\dj{}er Bo\v{s}kovi\'c Institute, 10000 Zagreb, Croatia} \\
$^{36}$ {INFN MAGIC Group: INFN Sezione di Perugia, 06123 Perugia, Italy} \\
$^{37}$ {INFN MAGIC Group: INFN Roma Tor Vergata, 00133 Roma, Italy} \\
$^{38}$ {now at University of Innsbruck} \\
$^{39}$ {also at Port d'Informaci\'o Cient\'ifica (PIC) E-08193 Bellaterra (Barcelona) Spain} \\
$^{40}$ {now at Ruhr-Universit\"at Bochum, Fakult\"at f\"ur Physik und Astronomie, Astronomisches Institut (AIRUB), 44801 Bochum, Germany} \\
$^{41}$ {also at Dipartimento di Fisica, Universit\`a di Trieste, I-34127 Trieste, Italy} \\
$^{42}$ {Max-Planck-Institut f\"ur Physik, D-80805 M\"unchen, Germany} \\
$^{43}$ {also at INAF Trieste and Dept. of Physics and Astronomy, University of Bologna} \\
$^{44}$ {Japanese MAGIC Group: Institute for Cosmic Ray Research (ICRR), The University of Tokyo, Kashiwa, 277-8582 Chiba, Japan} \\
$^{45}$ Dipartimento di Matematica e Fisica ``E. De Giorgi'', Universit\`a del Salento, Lecce, Italy \\
$^{46}$ Istituto Nazionale di Fisica Nucleare, Sezione di Lecce, I-73100 Lecce, Italy \\
$^{47}$ INAF-IRA Bologna, I-40129 Bologna, Italy \\
$^{48}$ Indian Institute of Astrophysics, Bangalore 560034, India \\
$^{49}$ Owens Valley Radio Observatory, California Institute of Technology, Pasadena, CA 91125, USA \\
$^{50}$ Finnish Center for Astronomy with ESO (FINCA), University of Turku, FI-20014, Turku, Finland \\ 
$^{51}$ Aalto University Mets\"ahovi Radio Observatory, Mets\"ahovintie 114, 02540 Kylm\"al\"a, Finland \\
$^{52}$ Institute of Astrophysics, Foundation for Research and Technology-Hellas, GR-71110 Heraklion, Greece \\
$^{53}$ Department of Physics, Univ. of Crete, GR-70013 Heraklion, Greece \\
$^{54}$ Departamento de Astronom\'ia, Universidad de Chile, Camino El Observatorio 1515, Las Condes, Santiago, Chile \\
$^{55}$ CePIA, Departamento de Astronom\'ia, Universidad de Concepti\'on, Concepci\'on, Chile \\
$^{56}$ Aalto University Department of Electronics and Nanoengineering, P.O. BOX 15500, FI-00076 AALTO, Finland.\\
$^{57}$ Mizusawa VLBI Observatory, National Astronomical Observatory of Japan, 2-12 Hoshigaoka, Mizusawa, Oshu, Iwate 023-0861, Japan \\
$^{58}$ Department of Astronomical Science, The Graduate University for Advanced Studies (SOKENDAI), 2-21-1 Osawa, Mitaka, Tokyo 181-8588, Japan \\
$^{59}$ Graduate School of Sciences and Technology for Innovation, Yamaguchi University, Yoshida 1677-1, Yamaguchi, Yamaguchi 753-8512, Japan \\
$^{60}$ Smithsonian Astrophysical Observatory, Cambridge, MA 02138, USA \\
$^{61}$ Astronomical Observatory, Jagiellonian University, ul. Orla 171, 30-244 Cracow, Poland


\appendix
\section{Search for hard GeV states}\label{sec:hardfermi}

\begin{figure}
   \centering
   \includegraphics[width=0.96\columnwidth]{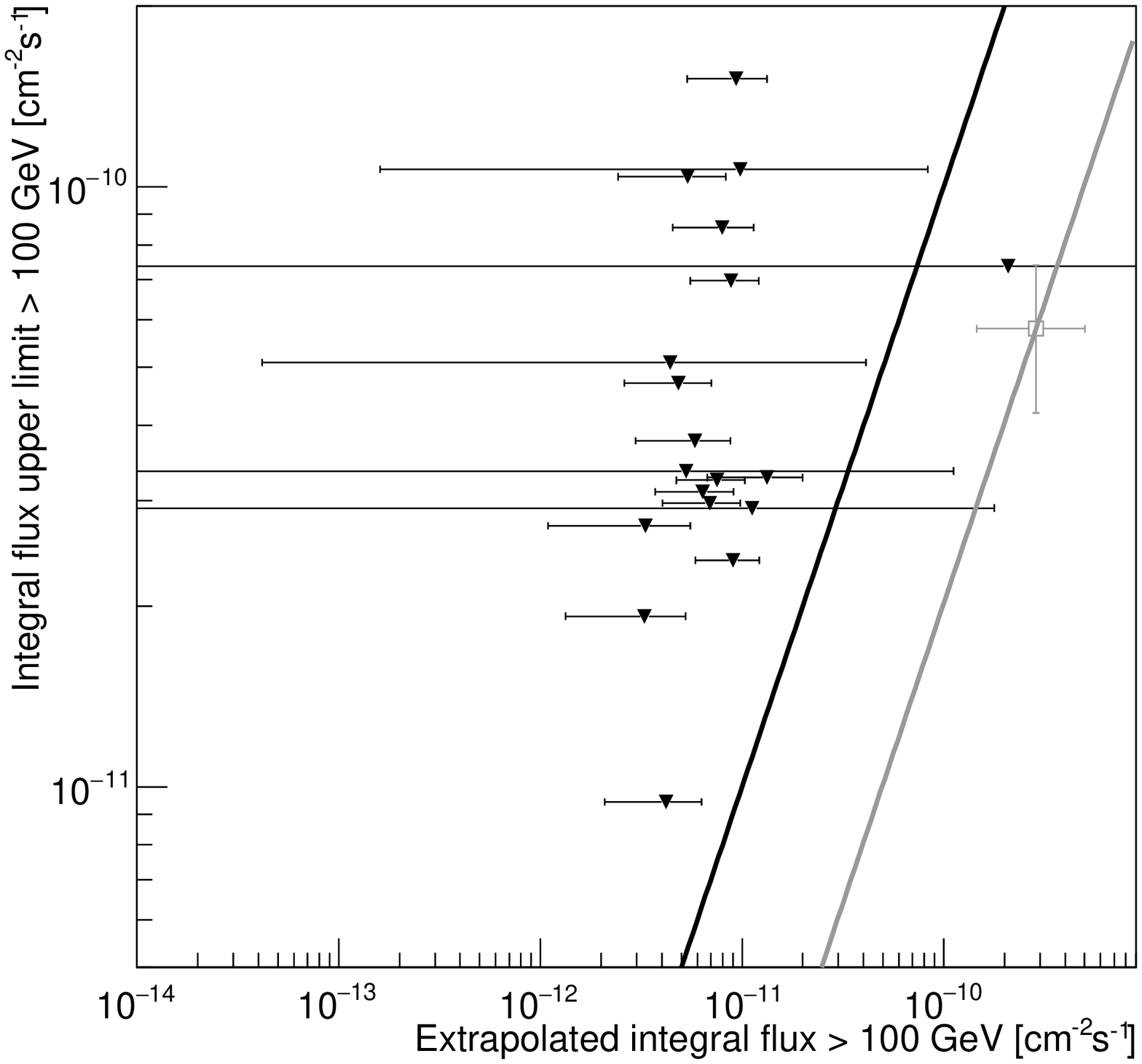}
   \caption{
     Integral upper limit on the flux $>100$\,GeV obtained with MAGIC telescopes as a function of the expected flux using contemporaneous \fermilat\ data (downward triangles).
     For comparison a measurement of the same quantities from the 2014 flare is shown in gray.
     Thick oblique lines show the proportionality of the two fluxes for the case when the true flux is equal to extrapolated one (black) or when the true flux scales with the expected one like in 2014 flare (gray)
   }  \label{fig:magic_extra}
\end{figure}

The \fermilat{} flux and photon index information are used to evaluate during which nights a detection with MAGIC would be most likely. 
The expected integral fluxes above 100 GeV were calculated  and compared with the obtained daily upper limits (see Fig.~\ref{fig:magic_extra}).
For each 24\,hr bin of \fermilat\ data that overlaps with MAGIC observations the \fermilat\ power-law spectrum were extrapolated to sub-TeV energies and convolved the flux with EBL absorption using the model of \citet{do11}.
Bins with \fermilat\ TS $< 9$ and those in which the uncertainty of the flux above 0.1\,GeV exceeds the flux value were removed from the analysis.
The VHE flux were integrated  and computed its uncertainty taking into account the uncertainty of the flux above 0.1\,GeV and the spectral index.
 It should be noted that in case of an intrinsic break or a cut-off in the spectrum the true flux will be lower than such extrapolated values.
 In fact, applying the same procedure to the 2014 flare data the measured flux is a factor of $\sim5$ below the extrapolated one.

In none of the bins with contemporaneous MAGIC observations the extrapolated flux value reached the flux of the 2014 flare, except for the case of MJD$=$58779 when such flux is consistent within the uncertainty bars. 

\section{Scenario for flares}\label{sec:flares}
In addition to the average emission, a few other interesting events occurred during the monitoring.
The MWL broadband SED during the X-ray flare (F2) shows a very similar shape in the synchrotron peak as during the average state. 
Despite higher X-ray flux, the GeV spectrum is consistent with the one obtained from the average state of the source.
Limited MWL data, unknown duration of X-ray flare, uncertain lensing and absorption (i.e. if the emission seen in different bands is dominantly from A or B image, that would affect magnification and absorption) and low statistics in gamma-ray data make modelling of this events difficult. 
In the framework of the model used for the explanation of the average emission the X-ray emission of the source is mainly of SSC origin.
Therefore, the event can be naturally explained by compression of the emission region, which would enhance the efficiency of this process. 
Such a compression (if it does not change the ambient magnetic field) would not modify synchrotron and EC components. 
Note also that a possible enhancement of the magnetic field during compression of the emission region does not have to increase considerably the synchrotron peak, as it is mainly explained in the average state modelling as the emission from large scale jet component.
Alternative explanation of the X-ray flare would involve enhanced emission from image A, which would show up in the hard part of the X-ray spectrum, while due to the strong absorption would not increase considerably the optical flux density. 
Unfortunately the X-ray data are not precise enough to allow us to distinguish between the two scenarios based on the X-ray absorption. 

The second interesting episode involves short optical flare during a longer GeV flare (F1). 
Hardening of the GeV spectrum and increase of the optical flux density points to hardening of the electron distribution and thus shifting of both peaks to higher energies.
The combination of different variability time scales in those two ranges makes the association of both events uncertain and complicates modelling of the emission. 
A possible scenario that would explain different time scales of optical and GeV emission would involve a blob travelling along the jet with a ramping up GeV emission. 
Since according to the low state modelling, most of the synchrotron emission is explained by ``Far'' emission zone (see Fig.~\ref{fig:mwl_model}) and thus such newly emerging blob would not show up as immediately enhanced optical emission. 
However if the new blob encounters a stationary feature in the jet, or an internal shock, it can cause enhancement of the magnetic field and shift of the synchrotron peak to higher energies. 
Since the SED of \srcs{} in optical range is very steep it would cause a strong optical flare, such as seen during period F1. 

The third investigated period, August 2020 MWL campaign cannot be firmly claimed as an enhanced flux state. 
Nevertheless the detection of two $>10$ GeV photons without accompanying clear increase of the flux at GeV energies, is consistent with a very hard electron energy distribution.
Unfortunately the VHE could be probed only in neighbouring nights. 
Curiously, a small hint of enhancement of the B-band flux is also consistent with hardening of the electron spectrum, as according to the low state model, the UV data probe the high energy part of the electron distribution. 
Short term wavelength-dependent variability in optical-UV  range could be then the effect of variability of the electron energy distribution convoluted with absorption in the lens galaxy. 
While no X-ray variability is present during August 2020, the average X-ray flux during this period is enhanced with respect to the low state, and is similar to the flux level of the F2 period. 
Within the framework of the modelling this could be explained if the compression of the emission region persisted between the MJD 58863.7 flare and August 2020. 
%


\bsp	
\label{lastpage}

\begin{thebibliography}{99}

\bibitem[Abdalla et al.(2019)]{ab19} H.~E.~S.~S. Collaboration, Abdalla, H., Aharonian, F., et al.\ 2019, \mnras, 486, 3886
\bibitem[Abdo et al.(2010)]{ssdc_1fgl} Abdo, A.~A., Ackermann, M., Ajello, M., et al.\ 2010, \apjs, 188, 405. doi:10.1088/0067-0049/188/2/405
\bibitem[\protect\citeauthoryear{Abdollahi, et al.}{2020}]{4fgl} Abdollahi S., et al., 2020, \apjs, 247, 33
\bibitem[Acciari et al.(2020)]{acc21} Acciari, V.~A., Ansoldi, S., Antonelli, L.~A., et al.\ 2020, arXiv:2012.11380 
\bibitem[Acero et al.(2015)]{ssdc_3fgl} Acero, F., Ackermann, M., Ajello, M., et al.\ 2015, \apjs, 218, 23. doi:10.1088/0067-0049/218/2/23
\bibitem[Acharya et al.(2013)]{2013APh....43....3A} Acharya, B.~S., Actis, M., Aghajani, T., et al.\ 2013, Astroparticle Physics, 43, 3. doi:10.1016/j.astropartphys.2013.01.007 
\bibitem[Ahnen et al.(2016)]{ah16} Ahnen, M.~L., Ansoldi, S., Antonelli, L.~A., et al.\ 2016, \aap, 595, A98 
\bibitem[Akaike(1974)]{ak74} Akaike, H., 1974, IEEE Trans. Autom. Control, AC-19, 716 
\bibitem[Aleksi{\'c} et al.(2016a)]{al16a} Aleksi{\'c}, J., Ansoldi, S., Antonelli, L.~A., et al.\ 2016, Astroparticle Physics, 72, 61 
\bibitem[Aleksi{\'c} et al.(2016b)]{al16b} Aleksi{\'c}, J., Ansoldi, S., Antonelli, L.~A., et al.\ 2016, Astroparticle Physics, 72, 76 
\bibitem[Algaba et al.(2015)]{algaba2015} Algaba, J.-C., Zhao, G.-Y., Lee, S.-S., et al. \ 2015, JKAS, 48, 237 
\bibitem[Arnaud(1996)]{arnaud96} Arnaud, K.~A.\ 1996, Astronomical Data Analysis Software and Systems V, 17
\bibitem[\protect\citeauthoryear{Atwood et al.}{2009}]{fermipaper} Atwood W.~B., Abdo A.~A., Ackermann M., Althouse W., Anderson B., Axelsson M., Baldini L., et al., 2009, ApJ, 697, 1071
\bibitem[Baars et al.(1977)]{1977A&A....61...99B} Baars, J.~W.~M., Genzel, R., Pauliny-Toth, I.~I.~K., et al.\ 1977, \aap, 500, 135 
\bibitem[Barnacka et al.(2011)]{ba11} Barnacka, A., Glicenstein, J.-F., \& Moudden, Y.\ 2011, \aap, 528, L3. doi:10.1051/0004-6361/201016175 
\bibitem[Barnacka et al.(2016)]{2016ApJ...821...58B} Barnacka, A., Geller, M.~J., Dell'Antonio, I.~P., et al.\ 2016, \apj, 821, 58. doi:10.3847/0004-637X/821/1/58
\bibitem[\protect\citeauthoryear{Barnacka et al.}{2015}]{2015ApJ...799...48B} Barnacka A., Geller M.~J., Dell'Antonio I.~P., Benbow W., 2015, ApJ, 799, 48. doi:10.1088/0004-637X/799/1/48
\bibitem[\protect\citeauthoryear{Barnacka}{2018}]{2018PhR...778....1B} Barnacka A., 2018, PhR, 778, 1. doi:10.1016/j.physrep.2018.10.001
\bibitem[van den Berg et al.(2019)]{be19} van den Berg, J.~P., B{\"o}ttcher, M., Dom{\'\i}nguez, A., et al.\ 2019, \apj, 874, 47 
\bibitem[Biggs et al.(1999)]{bi99} Biggs, A.~D., Browne, I.~W.~A., Helbig, P., et al.\ 1999, \mnras, 304, 349 
\bibitem[Biggs et al.(2003)]{bi03} Biggs, A.~D., Wucknitz, O., Porcas, R.~W., et al.\ 2003, \mnras, 338, 599. doi:10.1046/j.1365-8711.2003.06050.x 
\bibitem[Biggs \& Browne(2018)]{bb18} Biggs, A.~D., \& Browne, I.~W.~A.\ 2018, \mnras, 476, 5393 
\bibitem[Breeveld et al.(2010)]{breeveld10} Breeveld, A. A., et al. 2010, MNRAS, 406, 1687 
\bibitem[Burnham et al.(2011)]{bah11} Burnham, K.P., Anderson, D.R., \& Huyvaert, K.P., 2011, Behavioral Ecology and Sociobiology, 65, 23 
\bibitem[{{Burrows} {et~al.}(2004){Burrows}, {Hill}, {Nousek}, {Wells}, {Chincarini}, {Abbey}, {Beardmore}, {Bosworth}, {Br{\"a}uninger}, {Burkert}, {Campana}, {Capalbi}, {Chang}, {Citterio}, {Freyberg}, {Giommi}, {Hartner}, {Killough}, {Kittle}, {Klar}, {Mangels}, {McMeekin}, {Miles}, {Moretti},   {Mori}, {Morris}, {Mukerjee}, {Osborne}, {Short}, {Tagliaferri}, {Tamburelli}, {Watson}, {Willingale}, \&{Zugger}}]{2004SPIE.5165..201B} {Burrows}, D.~N., {Hill}, J.~E., {Nousek}, J.~A., {et~al.} 2004, in \procspie, Vol. 5165, X-Ray and Gamma-Ray Instrumentation for Astronomy XIII, ed. K.~A. {Flanagan} \& O.~H.~W. {Siegmund}, 201--216 
\bibitem[Buson et al.(2015a)]{2015arXiv150203134B} Buson, S., Cheung, C.~C., Larsson, S., et al.\ 2015, arXiv:1502.03134 
\bibitem[Buson et al.(2015b)]{2015ICRC...34..877B} Buson, S., Cheung, C.~T., Larsson, S., et al.\ 2015, 34th International Cosmic Ray Conference (ICRC2015), 34, 877
\bibitem[Carilli et al.(1993)]{cry93} Carilli, C.~L., Rupen, M.~P., \& Yanny, B.\ 1993, \apjl, 412, L59 
\bibitem[\protect\citeauthoryear{Cash}{1979}]{1979ApJ...228..939C} Cash W., 1979, ApJ, 228, 939 
\bibitem[Ceribella et al.(2019)]{ce19} Ceribella, G., D'Amico, G., Dazzi, F., et al.\ 2019, 36th International Cosmic Ray Conference (ICRC2019), 645 
\bibitem[Cheung et al.(2014)]{ch14} Cheung, C.~C., Larsson, S., Scargle, J.~D., et al.\ 2014, \apjl, 782, L14 
\bibitem[Cohen et al.(2000)]{co00} Cohen, A.~S., Hewitt, J.~N., Moore, C.~B., \& Haarsma, D.~B.\ 2000, \apj, 545, 578
\bibitem[Cohen et al.(2003)]{co03} Cohen, J.~G., Lawrence, C.~R., \& Blandford, R.~D.\ 2003, \apj, 583, 67 
\bibitem[Condon et al.(1998)]{ssdc_nvss} Condon, J.~J., Cotton, W.~D., Greisen, E.~W., et al.\ 1998, \aj, 115, 1693. doi:10.1086/300337
\bibitem[Corbett et al.(1996)]{co96} Corbett, E.~A., Browne, I.~W.~A., Wilkinson, P.~N., \& Patnaik, A.\ 1996, Astrophysical Applications of Gravitational Lensing, 173, 37 
\bibitem[Costamante et al.(2018)]{co18} Costamante, L., Cutini, S., Tosti, G., et al.\ 2018, \mnras, 477, 4749 
\bibitem[Dazzi et al.(2021)]{da21} Dazzi., F. et al., IEEE Transactions on Nuclear Science, doi: 10.1109/TNS.2021.3079262.
\bibitem[Dermer \& Menon(2009)]{dm09} Dermer, C.~D. \& Menon, G.\ 2009, High Energy Radiation from Black Holes: Gamma Rays, Cosmic Rays, and Neutrinos by Charles D. Dermer and Govind Menon. Princeton Univerisity Press, November 2009.
\bibitem[D'Elia et al.(2013)]{ssdc_1swxrt} D'Elia, V., Perri, M., Puccetti, S., et al.\ 2013, \aap, 551, A142. doi:10.1051/0004-6361/201220863
\bibitem[Dom{\'{\i}}nguez et al.(2011)]{do11} Dom{\'{\i}}nguez, A., Primack, J.~R., Rosario, D.~J., et al.\ 2011, \mnras, 410, 2556 
\bibitem[Eulaers \& Magain(2011)]{em11} Eulaers, E., \& Magain, P.\ 2011, \aap, 536, A44 
\bibitem[{{Evans} {et~al.}(2009){Evans}, {Beardmore}, {Page}, {Osborne}, {O'Brien}, {Willingale}, {Starling}, {Burrows}, {Godet}, {Vetere}, {Racusin}, {Goad}, {Wiersema}, {Angelini}, {Capalbi}, {Chincarini}, {Gehrels}, {Kennea}, {Margutti}, {Morris}, {Mountford}, {Pagani}, {Perri}, {Romano}, \& {Tanvir}}]{2009MNRAS.397.1177E} {Evans}, P.~A., {Beardmore}, A.~P., {Page}, K.~L., {et~al.} 2009, \mnras, 397, 1177 
\bibitem[Evans et al.(2014)]{ssdc_1sxps}  Evans, P. A.; Osborne, J. P.; Beardmore, A. P. et al., Astrophys. J. Suppl. Ser., 210, 8 
\bibitem[Falco et al.(1999)]{fa99} Falco, E.~E., Impey, C.~D., Kochanek, C.~S., et al.\ 1999, \apj, 523, 617 
\bibitem[{{Fallah Ramazani} {et~al.}(2017){Fallah Ramazani}, {Lindfors}, \&  {Nilsson}}]{2017A&A...608A..68F} {Fallah Ramazani}, V., {Lindfors}, E., \& {Nilsson}, K. 2017, \aap, 608, A68 
\bibitem[Falomo et al.(2017)]{fa17} Falomo, R., Treves, A., Scarpa, R., et al.\ 2017, \mnras, 470, 2814 
\bibitem[Finke(2016)]{fi16} Finke, J.~D.\ 2016, \apj, 830, 94
\bibitem[Fruck \& Gaug(2015)]{fg15} Fruck, C., \& Gaug, M.\ 2015, European Physical Journal Web of Conferences, 89, 02003 
\bibitem[Gabriel et al.(2004)]{gabriel04} Gabriel, C., Denby, M., Fyfe, D.~J., et al.\ 2004, Astronomical Data Analysis Software and Systems (ADASS) XIII, 759
\bibitem[Ghisellini \& Tavecchio(2009)]{2009MNRAS.397..985G} Ghisellini, G. \& Tavecchio, F.\ 2009, \mnras, 397, 985. doi:10.1111/j.1365-2966.2009.15007.x
\bibitem[Ghisellini et al.(2010)]{gh10} Ghisellini, G., Tavecchio, F., Foschini, L., et al.\ 2010, \mnras, 402, 497
\bibitem[Gregory et al.(1996)]{ssdc_gb6} Gregory, P.~C., Scott, W.~K., Douglas, K., et al.\ 1996, \apjs, 103, 427. doi:10.1086/192282
\bibitem[Gregory \& Condon(1991)]{ssdc_gb87} Gregory, P.~C. \& Condon, J.~J.\ 1991, \apjs, 75, 1011. doi:10.1086/191559
\bibitem[Greisen(2003)]{greisen2003} Greisen, E.~W. \ 2003, ASSL, 28, 71
\bibitem[\protect\citeauthoryear{Gottlieb}{1994}]{1994JMP....35.5507G} Gottlieb D.~H., 1994, JMP, 35, 5507. doi:10.1063/1.530762
\bibitem[Hada et al.(2017)]{hada2017} Hada, K., Park, J., Kino, M., et al. \ 2017, PASJ, 69, 71
\bibitem[Hada et al.(2020)]{hada2020} Hada, K., Niinuma, K., Sitarek, J., et al.\ 2020, \apj, 901, 2. doi:10.3847/1538-4357/abaab1
\bibitem[Jackson, Xanthopoulos \& Browne (2000)]{jackson00} Jackson, N., Xanthopoulos, E., Browne, I.W.A., 2000, \mnras, 311, 389 
\bibitem[Jackson et al.(2007)]{ssdc_jvaspol} Jackson, N., Battye, R.~A., Browne, I.~W.~A., et al.\ 2007, \mnras, 376, 371. doi:10.1111/j.1365-2966.2007.11442.x
\bibitem[Jansen et al.(2001)]{jansen01} Jansen, F., Lumb, D., Altieri, B., et al.\ 2001, \aap, 365, L1
\bibitem[Kalberla et al.(2005)]{Kalberla05} Kalberla, P.~M.~W., Burton, W.~B., Hartmann, D., et al.\ 2005, A\&A, 440, 775
\bibitem[Kuehr et al.(1981)]{ssdc_kuehr} Kuehr, H., Witzel, A., Pauliny-Toth, I.~I.~K., et al.\ 1981, \aaps, 45, 367
\bibitem[\protect\citeauthoryear{Keeton}{2001}]{2001astro.ph..2341K} Keeton C.~R., 2001, arXiv, astro-ph/0102341
\bibitem[Larchenkova et al.(2011)]{2011AstL...37..233L} Larchenkova, T.~I., Lutovinov, A.~A., \& Lyskova, N.~S.\ 2011, Astronomy Letters, 37, 233. doi:10.1134/S1063773711040050
\bibitem[Lee et al.(2015)]{lee2015} Lee, S.-S.,  Oh, C.-S.,  Roh, D.-G., et al. \ 2015, JKAS, 48, 125
\bibitem[MAGIC Collaboration et al.(2020)]{acc20} MAGIC Collaboration, Acciari, V.~A., Ansoldi, S., et al.\ 2020, \aap, 640, A132. doi:10.1051/0004-6361/202037811
\bibitem[Mannucci et al. (2001)]{mannu01} Mannucci F., Basile, F., Poggianti, B.~M., Cimatti, A., Daddi, E., Pozzetti, L., Vanzi, L., 2001, \mnras, 326, 745
\bibitem[Menten \& Reid(1996)]{mr96} Menten, K.~M., \& Reid, M.~J.\ 1996, \apjl, 465, L99
\bibitem[Mittal et al.(2006)]{mi06} Mittal, R., Porcas, R., Wucknitz, O., Biggs, A., \& Browne, I.\ 2006, A\&A, 447, 515
\bibitem[Mittal et al.(2007)]{mi07} Mittal, R., Porcas, R., \& Wucknitz, O.\ 2007, A\& A, 465, 405
\bibitem[Myers et al.(2003)]{ssdc_class} Myers, S.~T., Jackson, N.~J., Browne, I.~W.~A., et al.\ 2003, \mnras, 341, 1. doi:10.1046/j.1365-8711.2003.06256.x
\bibitem[Nieppola et al.(2007)]{ssdc_nieppo} Nieppola, E., Tornikoski, M., L{\"a}hteenm{\"a}ki, A., et al.\ 2007, \aj, 133, 1947. doi:10.1086/512609 
\bibitem[Nigro et al.(2020)]{agnpy} Nigro, C., Sitarek, J.,  Craig M. \& Gliwny, P. (2020, September 28). agnpy: modelling Active Galactic Nuclei radiative processes with python. Zenodo. http://doi.org/10.5281/zenodo.4055176  
\bibitem[Niinuma et al.(2014)]{niinuma2014} Niinuma, K., Lee, S.-S., Kino, M., et al. \ 2014, PASJ, 66, 103 
\bibitem[Nilsson et al.(2018)]{nilsson18} Nilsson, K., Lindfors, E., Takalo, L.O., et al. \ 2018, \aap, 620, 185
\bibitem[Nolan et al.(2012)]{ssdc_2fgl} Nolan, P.~L., Abdo, A.~A., Ackermann, M., et al.\ 2012, \apjs, 199, 31. doi:10.1088/0067-0049/199/2/31
\bibitem[O'Dea et al.(1992)]{od92} O'Dea, C.~P., Baum, S.~A., Stanghellini, C., et al.\ 1992, \aj, 104, 1320 
\bibitem[Olgu\'{i}n-Iglesias et al. (2016)]{olguin16} Olgu\'{i}n-Iglesias, A., Le\'{o}n-Tavares, J., Kotilainen, J. K., et al. \ 2016, \mnras, 460, 3202
\bibitem[Paiano et al.(2017)]{pa17} Paiano, S., Landoni, M., Falomo, R., et al.\ 2017, \apj, 837, 144  
\bibitem[Paliya et al.(2021)]{2021ApJS..253...46P} Paliya, V.~S., Dom{\'\i}nguez, A., Ajello, M., et al.\ 2021, \apjs, 253, 46. doi:10.3847/1538-4365/abe135
\bibitem[Patnaik, Porcas, \& Browne (1995)]{1995MNRAS.274L...5P} Patnaik A.~R., Porcas R.~W., Browne I.~W.~A., 1995, MNRAS, 274, L5. 
\bibitem[\protect\citeauthoryear{Petters \& Werner}{2010}]{2010GReGr..42.2011P} Petters A.~O., Werner M.~C., 2010, GReGr, 42, 2011. doi:10.1007/s10714-010-0968-6
\bibitem[Planck Collaboration et al.(2011)]{ssdc_ercsc} Planck Collaboration, Ade, P.~A.~R., Aghanim, N., et al.\ 2011, \aap, 536, A7. doi:10.1051/0004-6361/201116474
\bibitem[Planck Collaboration et al.(2014)]{ssdc_pccs1} Planck Collaboration, Ade, P.~A.~R., Aghanim, N., et al.\ 2014, \aap, 571, A28. doi:10.1051/0004-6361/201321524
\bibitem[Planck Collaboration et al.(2016)]{ssdc_pccs2} Planck Collaboration, Ade, P.~A.~R., Aghanim, N., et al.\ 2016, \aap, 594, A26. doi:10.1051/0004-6361/201526914
\bibitem[Poole et al.(2008)]{poole08} Poole, T. S., et al. 2008, MNRAS, 383, 627 
\bibitem[Readhead et al.(1989)]{1989ApJ...346..566R} Readhead, A.~C.~S., Lawrence, C.~R., Myers, S.~T., et al.\ 1989, \apj, 346, 566. doi:10.1086/168039
\bibitem[Richards et al.(2011)]{2011ApJS..194...29R} Richards, J.~L., Max-Moerbeck, W., Pavlidou, V., et al.\ 2011, \apjs, 194, 29. doi:10.1088/0067-0049/194/2/29 
\bibitem[Rolke et al.(2005)]{ro05} Rolke, W.~A., L{\'o}pez, A.~M., \& Conrad, J.\ 2005, NIM A, 551, 493 
\bibitem[Schlafly \& Finkbeiner(2011)]{sch11} Schlafly, E.~F. \& Finkbeiner, D.~P.\ 2011, \apj, 737, 103. doi:10.1088/0004-637X/737/2/103
\bibitem[Shayduk(2013)]{sh13} Shayduk, M., \ 2013, 33rd International Cosmic Ray Conference, 3000, astro-ph.IM/1307.4939
\bibitem[Shepherd et al.(1994)]{shepherd1994} Shepherd, M.~C., Pearson, T.~J., \& Taylor, G.~B. \ 1994, BAAS, 26, 987
\bibitem[Sitarek \& Bednarek(2016)]{sb16} Sitarek, J., \& Bednarek, W.\ 2016, \mnras, 459, 1959 
\bibitem[Str{\"u}der et al.(2001)]{stru01} Str{\"u}der, L., Briel, U., Dennerl, K., et al.\ 2001, \aap, 365, L18
\bibitem[Ter\"asranta et al. (1998)]{te98} Ter\"asranta, H., Tornikoski, M., Mujunen, A. et al. 1998, A\&AS, 132, 305 
\bibitem[Vovk \& Neronov(2016)]{vn15} Vovk, I., \& Neronov, A.\ 2016, \aap, 586, A150
\bibitem[Wilms et al.(2000)]{Wilms00} Wilms, J., Allen, A., \& McCray, R.\ 2000, \apj, 542, 914
\bibitem[\protect\citeauthoryear{Wood, et al.}{2017}]{fermipy} Wood M., Caputo R., Charles E., Di Mauro M., Magill J., Perkins J.~S., Fermi-LAT Collaboration, 2017, ICRC, 301, 824, ICRC...35
\bibitem[Wright et al.(2009)]{ssdc_wmap5} Wright, E.~L., Chen, X., Odegard, N., et al.\ 2009, \apjs, 180, 283. doi:10.1088/0067-0049/180/2/283
\bibitem[Wright et al.(2010)]{ssdc_allwise} Wright, E.~L., Eisenhardt, P.~R.~M., Mainzer, A.~K., et al.\ 2010, \aj, 140, 1868. doi:10.1088/0004-6256/140/6/1868
\bibitem[Wong et al. (2020)]{2020MNRAS.498.1420W} Wong K.~C., Suyu S.~H., Chen G.~C.-F., Rusu C.~E., Millon M., Sluse D., Bonvin V., et al., 2020, \mnras, 498, 1420.
\bibitem[Wucknitz et al.(2004)]{2004MNRAS.349...14W} Wucknitz, O., Biggs, A.~D., \& Browne, I.~W.~A.\ 2004, \mnras, 349, 14. doi:10.1111/j.1365-2966.2004.07514.x
\bibitem[York et al.(2005)]{2005MNRAS.357..124Y} York, T., Jackson, N., Browne, I.~W.~A., et al.\ 2005, \mnras, 357, 124. doi:10.1111/j.1365-2966.2004.08618.x
\bibitem[Zanin et al.(2013)]{za13} Zanin, R., Carmona, E., Sitarek, J., et al., 2013, Proc of 33rd ICRC, Rio de Janeiro, Brazil, Id. 773 
\bibitem[\protect\citeauthoryear{Zhang et al.}{2007}]{2007MNRAS.377.1623Z} Zhang M., Jackson N., Porcas R.~W., Browne I.~W.~A., 2007, MNRAS, 377, 1623. doi:10.1111/j.1365-2966.2007.11718.x
\bibitem[Zhao et al.(2019)]{zhao2019} Zhao, G.-Y., Jung, T., Sohn, B.~W., et al. \ 2019, JKAS, 52, 23   
\end{thebibliography}
\end{document}